\documentclass[a4paper,12pt]{article}
\pdfoutput=1
\usepackage{jheppub}
\usepackage{graphicx}
\usepackage{amssymb}
\usepackage{amsmath}
\usepackage{amsfonts}
\usepackage{hyperref}
\usepackage{xcolor}

	\def\bea{\begin{eqnarray}}
	\def\eea{\end{eqnarray}}
	\def\be{\begin{equation}}
	\def\ee{\end{equation}}
	\def\nn{\nonumber\\}

	\def\mm{\mathcal}

	\definecolor{db}{rgb}{0,0.08,0.45}
	\definecolor{brick}{rgb}{0.6,0.1,0.3}
	
	\definecolor{zz}{rgb}{1,0,0}
	\definecolor{zz2}{rgb}{0.7,0.1,0.1}
	\definecolor{yy}{rgb}{0.05,0.9,0.05}
	\definecolor{ww}{rgb}{0.6,0.1,0.3}
	
	\definecolor{rr}{cmyk}{0,0,0,1}
	\definecolor{vv}{rgb}{0.5,0,0.5}
	\definecolor{ss}{cmyk}{0,0,0,1}

	\definecolor{brick}{rgb}{0.5,0,0.5}

	\def\a{\alpha}       \def\n{\nu} \def\m{\mu}     \def\l{\lambda}  
	\def\G{\Gamma}  \def\D{\Delta}

\title{1d Conformal Field Theory and Dispersion Relations}

\author[a]{Dean Carmi,}
\affiliation[a]{Department of Mathematics and Physics University of Haifa at Oranim, Kiryat Tivon 36006, Israel}
\author[b,c]{Sudip Ghosh,}
\author[b,c]{Trakshu Sharma}
\affiliation[b]{Department of Physics and Haifa Center for Physics and Astrophysics,
University of Haifa, Haifa 3498838, Israel}
\affiliation[c]{Department of Physics, Technion, Israel Institute of Technology, Haifa, 32000, Israel}
\emailAdd{deancarmi1@gmail.com, sudip112phys@gmail.com, trakshusharma@gmail.com
 }

\abstract{We study conformal field theory in $d=1$ space-time dimensions. We derive a dispersion relation for the 4-point correlation function of identical bosons and fermions, in terms of the double discontinuity. This extends the conformal dispersion relation of \cite{Carmi:2019cub}, which holds for CFTs in dimensions $d\geq 2$, to the case of $d=1$. The dispersion relation is obtained by combining the Lorentzian inversion formula with the operator product expansion of the 4-point correlator. We perform checks of the dispersion relation using correlators of generalised free fields and derive an integral relation between the kernel of the dispersion relation and that of the Lorentzian inversion formula. Finally, for $1$-$d$ holographic conformal theories, we analytically compute scalar Witten diagrams in $AdS_2$ at tree-level and $1$-loop.

}

\begin{document} 
\maketitle
\flushbottom



\newpage
\section{Introduction}
\label{sec:intro}

Quantum field theories that have scale and conformal symmetry, i.e. conformal field theories (CFT) find applications in diverse areas of physics such as in the study of critical phenomena, statistical mechanics and string theory. A powerful non-perturbative tool for solving CFTs is the conformal bootstrap, which leverages crossing symmetry, unitarity and the operator product expansion (OPE) to constrain and determine the spectrum, labelled by the scaling dimensions $\Delta_{\phi}$ and spin $\ell$ of local operators, and the OPE coefficients of the CFT. In recent years, several numerical \cite{Poland:2018epd,Poland:2022qrs,Rychkov:2023wsd} as well as analytical \cite{Hartman:2022zik,Bissi:2022mrs} techniques have been developed to implement the conformal bootstrap for carving out the space of consistent CFTs.

A key recent advancement in the analytic bootstrap domain has been the introduction of the Lorentzian inversion formula (LIF) and dispersion relations for CFTs. The LIF, derived in \cite{Caron-Huot:2017vep}, extracts the OPE data from an integral transform of the double discontinuity of a four-point CFT correlator in dimensions $d\ge 2$. In \cite{Carmi:2019cub}, it was shown that the LIF in turn implies the existence of a dispersion relation which reconstructs the CFT $4$-point function from its double discontinuity in two OPE channels. The utility of the double discontinuity appearing in the inversion formula and the dispersion relation lies in its positivity properties and the fact that it can be efficiently computed in many cases. The LIF and consequently the conformal dispersion relation also encode the action of certain functionals which give rise to sum rules for the CFT's OPE coefficients.  Dispersive sum rules have been applied to derive bounds on the CFT data \cite{Mazac:2016qev,Mazac:2018mdx,Mazac:2019shk,Mazac:2018ycv,Caron-Huot:2020adz,Carmi:2020ekr,Knop:2022viy}. 

The dispersion relation of \cite{Carmi:2019cub} manifests crossing symmetry of the CFT four-point correlator in two out of three OPE channels. Dispersive representations which exhibit full crossing symmetry have been recently derived for Mellin amplitudes in \cite{Gopakumar:2021dvg} and for position space correlators in \cite{Bissi:2022fmj}. However, these dispersion relations involve the single discontinuity rather than the double discontinuity. Besides the conformal bootstrap, dispersion relations are also an important tool in the context of the S-matrix bootstrap program, where they have been used for example to obtain two-sided bounds on Wilson coefficients in the low-energy expansion of $2$-to-$2$ scattering amplitudes \cite{Kruczenski:2022lot,Caron-Huot:2020cmc,Caron-Huot:2021rmr,Sinha:2020win,Chowdhury:2021ynh,deRham:2022hpx}.

In this paper, we focus on the study of one-dimensional conformal field theories. There exist several interesting systems featuring one-dimensional conformal symmetry such as line defects in the $3$-d Ising model \cite{Billo:2013jda,Gaiotto:2013nva}, defect $1$-$d$ CFTs on Wilson lines in $\mathcal{N}=4$ super Yang-Mills theory \cite{Giombi:2017cqn,Beccaria:2017rbe,Giombi:2018qox,Liendo:2018ukf,Giombi:2018hsx,Gimenez-Grau:2019hez,Beccaria:2019dws} and ABJM theory \cite{Bianchi:2017svd,Bianchi:2017ozk,Bianchi:2018scb,Drukker:2019bev,Bianchi:2020hsz,Penati:2021tfj,Castiglioni:2022yes,Castiglioni:2023uus,Castiglioni:2023tci}, and SYK models \cite{Maldacena:2016hyu,Gross:2017vhb}. The four-point correlation function in this case depends on only a single cross-ratio $z$, unlike two cross-ratos $(z,\bar{z})$ in higher dimensions and hence it can be regarded as the restriction of the four-point function in a $d\ge 2$ CFT to the kinematic configuration $z=\bar{z}$. Recently, via methods such as the functional approach \cite{Mazac:2016qev,Mazac:2018mdx,Mazac:2018ycv,Paulos:2019fkw,Ghosh:2023wjn} to the conformal bootstrap and the $1$-$d$ Polyakov-Mellin bootstrap \cite{Ferrero:2019luz}, significant progress has been achieved in the analytic exploration of the crossing symmetry equation, which is simpler in the $1$-$d$ case.

A central result of this paper is the derivation of a $1$-$d$ CFT dispersion relation for the four-point function of identical operators, which can be either bosons or fermions. The main input in this derivation is the LIF of \cite{Mazac:2018qmi} which is applicable for $1$-$d$ CFTs, unlike the LIF of \cite{Caron-Huot:2017vep} which only works for $d\ge 2$. Similar to the higher dimensional case of \cite{Carmi:2019cub}, this $1$-$d$ dispersion relation expresses the four-point correlation function as an integral of its double discontinuity times a kernel. The main feature of this dispersion relation which distinguishes it from the higher-dimensional counterpart in \cite{Carmi:2019cub} is that it is manifestly crossing symmetric. The kernel in this dispersion relation also explicitly depends on the scaling dimension $\Delta_{\phi}$ of the external operator $\phi(x)$ in the four-point function, which is a consequence of the dispersive representation being crossing symmetric. The construction of dispersion relations for $1$-$d$ CFT four-point functions has also recently appeared in \cite{Paulos:2020zxx, Bonomi:2024lky}. Conformal dispersion relations for defect and boundary CFTs have been obtained in \cite{Bianchi:2022ppi,Barrat:2022psm}.

\subsection*{Outline and Summary}

In Section \ref{sec:dg2} we review the derivation of the dispersion relation of \cite{Carmi:2019cub} for CFTs in $d\ge 2$ dimensions.

In Section \ref{sec:deq1}, we use the Lorentzian OPE inversion formula of \cite{Mazac:2018qmi} to derive a dispersion relation for the $4$-point function of identical operators in 1d CFT. The kernel in the dispersion relation contains two types of terms which come from the LIF for the principal and discrete series representations of $SL(2,\mathbb{R})$ respectively. The contribution from the discrete series is independent of the external operator's scaling dimensions $\Delta_{\phi}$. We evaluate this in Section \ref{Discrete_Kernel}. The term corresponding to the principal series however depends on $\Delta_{\phi}$. We can compute it explicitly only when $\Delta_{\phi}$ takes either an integer value in the bosonic case or a half-integer value in the fermionic case. We consider the case $\Delta_{\phi}\in \mathbb{N}$ in Section \ref{subsec:KBprinc}. The dispersion kernel derived in this section however only works for correlators which are super-bounded in the Regge limit. In Section \ref{subsec:KBimproved}, we obtain an improved dispersion relation which can be applied to bosonic correlators that satisfy a weaker Regge bound. The evaluation of the dispersion kernel in the fermionic case for  $\Delta_{\phi}\in \mathbb{N}+1/2$ is discussed in Section \ref{subsec:KFprinc}.  In Section \ref{sec:Pauloscomp}, we comment on the difference between our method for deriving the dispersion relation and the method of \cite{Paulos:2020zxx}, which uses the Polyakov bootstrap. 

Section \ref{sec:checks} deals with checks of the dispersion relation using generalised free theory correlators for $\Delta_{\phi}=1$ and $\Delta_{\phi}=1/2$. 

Section \ref{sec:HandKrel} contains the derivation of an integral relation between the dispersion kernel and the Lorentzian OPE inversion kernel, which arises by plugging the dispersion relation in the Euclidean OPE inversion formula.

In Section \ref{sec:5}, we consider the computation of Witten diagrams at tree-level as well as at $1$-loop for a scalar field theory in $AdS_{2}$ with cubic and quartic interaction terms.  

We conclude with some directions for future work in Section \ref{sec:concl}.  The appendices provide further details on some of the computations in this paper. In Sections \ref{K_princ_p_Appen} and \ref{app:Kqprinc} we expand on the derivation of the principal series part of the dispersion kernel. Section \ref{sums} contains a list of sums involving LegendreP's and LegendreQ's which are useful in the explicit evaluation of the dispersion kernel.  \\

\vskip 4pt
\noindent \textbf{Note}:
While finishing our work, the paper \cite{Bonomi:2024lky} appeared on Arxiv. This paper also derives a dispersion relation for the $4$-point correlator of identical bosonic and fermionic operators in $1$-$d$ CFTs, using the LIF of \cite{Mazac:2018qmi}. In \cite{Bonomi:2024lky}, the $1$-$d$ dispersion relation has also been used to compute perturbative corrections to generalised free theory correlators coming from scalar quartic contact interaction in $AdS_{2}$ and for the evaluation of correlators of the defect CFT$_{1}$ on the half-BPS Wilson line in $\mathcal{N}=4$ super Yang-Mills theory, perturbatively around large 't Hooft coupling $\lambda$. Our results are in agreement with those of \cite{Bonomi:2024lky} wherever there is overlap.


\subsection{Conformal dispersion relation for $d \geq 2$}
\label{sec:dg2}

A dispersion relation is a formula that gives a function in terms of an integral over its discontinuities. The simplest type of dispersion relations are single-variable Cauchy-type dispersion relations. For example, the 2-to-2 relativistic scattering amplitude $\mm{M}(s,t)$ can be shown to obey a Cauchy-type dispersion relation\footnote{For a Cauchy-type dispersion relation in the context of CFT 4-point correlators, see \cite{Bissi:2019kkx}.}
\begin{equation}
\label{eq:ds4}
\mm{M}(s,t) = \frac{1}{2\pi} \int \frac{Disc_{s'}[\mm{M}(s',t)]}{s'-s} +(s\leftrightarrow u)
\end{equation}
where the integral is along the branch cuts of the amplitude $\mm M$, and the (single) discontinuity across the cut is:
\begin{equation}
i Disc_{s'}[\mm{M}(s',t)] \equiv  \mm{M}(s'+i 0,t) - \mm{M}(s'-i 0,t) \ .
\end{equation}
The standard way to derive the Cauchy-type dispersion relation of Eq.~\eqref{eq:ds4} is via a contour deformation argument. Specifically, one starts by using Cauchy's residue theorem for the function $\frac{\mm{M}(s',t)}{s'-s}$, then the contour is deformed to wrap the branch cuts, which results in the dispersion relation Eq.~\eqref{eq:ds4}. For the derivation to work, the amplitude $\mm{M}(s,t)$ has to be an analytic function apart from cuts, and it must decay fast enough as $s\to \infty$. 

An alternative derivation of the dispersion relation starts by assuming the Froissart-Gribov inversion formula and inserting it in the partial-wave expansion of $\mm{M}(s,t)$. For a review of this procedure we refer the reader to section 2.1 of \cite{Carmi:2019cub}. Instead, we will now review details of the derivation of the conformal dispersion relation, which has a similar essence.

A conformal dispersion relation for the 4-point correlation functions of a CFT${}_{d\geq 2}$, was derived in \cite{Carmi:2019cub}. This formula gives the 4-point correlator $\mm{G}(z,\bar z)$ in terms of an integral over its double discontinuity. This formula was derived by combining the Lorentzian inversion formula and the OPE expansion of the correlator $\mm{G}(z,\bar z)$. We now summarize the main steps of this derivation. The Lorentzian inversion formula of \cite{Caron-Huot:2017vep} is:
\begin{equation}
\label{eq:fd1}
c(J,\D) =  c^t(J,\D)+(-1)^J c^u(J,\D)
\end{equation}
where
\begin{equation}
\label{eq:fd2}
c^t(J,\D) =\frac{\kappa_{J+\D}}{4}\int_0^1 dw d\bar w \m(w, \bar w) G_{\D+1-d,J+d-1}(w, \bar w) \mathrm{dDisc}[\mm{G}(w,\bar w)].
\end{equation}
Here $c^t(J,\D)$ is the $t$-channel OPE function, $G_{\D+1-d,J+d-1}(w, \bar w)$ is the ``inverted" conformal block, and $ \mathrm{dDisc}[\mm{G}(z,\bar z)]$ is the double discontinuity around $\bar w=1$, defined as:
\begin{equation}
 \mathrm{dDisc}[\mm{G}(z,\bar z)] \equiv \cos(\pi(a+b)) \mm{G}(z,\bar z)-  \frac{1}{2}e^{i\pi(a+b)}\Big( \mm{G}(z,\bar z^\circlearrowleft) +\mm{G}(z,\bar z^\circlearrowright) \Big).
\end{equation}
The conformal partial wave expansion of $\mm{G}(z,\bar z)$ can be written as:
\begin{equation}
\label{eq:fd3}
\mm{G}(z,\bar z)= \sum_{J=0}^\infty \int_{\frac{d}{2}-i\infty}^{\frac{d}{2}+i\infty}  \frac{d\D}{2\pi i} \ c(J,\D) F_{J,\D}(z,\bar z)
\end{equation}
where $F_{J,\D}(z,\bar z)$ is the conformal partial wave, defined as the combination of the conformal block and its shadow\footnote{The coefficient of the shadow block is not important for our purposes.}:
\begin{equation}
F_{J,\D}(z, \bar z)= \frac{1}{2}\Big( G_{J,\D}(z,\bar z) + \# G_{J,d-\D}(z,\bar z) \Big).
\end{equation}
To derive the conformal dispersion relation, we insert the Lorentzian inversion formula Eq.~\eqref{eq:fd1} and \eqref{eq:fd2} inside the conformal partial wave expansion Eq.~\eqref{eq:fd3}. The result is:
\begin{equation}
\mm{G}^t(z,\bar z) = \int_0^1 dw d\bar w \ K(z,\bar z,w,\bar w)  \mathrm{dDisc}[\mm{G}(w,\bar w)]
\end{equation}
where the kernel is:
\begin{equation}
K(z,\bar z,w,\bar w)= \frac{\m(w, \bar w)}{8\pi i} \sum_{J=0}^\infty \int_{\frac{d}{2}-i\infty}^{\frac{d}{2}+i\infty} d\D \ \kappa_{J+\D}F_{J,\D}(z, \bar z)G_{\D+1-d,J+d-1}(w, \bar w).
\end{equation}
The full correlator is the sum of the $t$ and $u$-channel contributions:
\begin{equation}
\mm{G}(z,\bar z) = \mm{G}^t(z,\bar z) +\mm{G}^u(z,\bar z) \ .
\end{equation}
The dispersion kernel $K(z,\bar z,w,\bar w)$ is a function of 4 cross-ratios, and a major goal of \cite{Carmi:2019cub} was to compute it analytically in terms of these variables. The result is that $K$ can be written as a hypergeometric ${}_2F_1$ function, and we refer to \cite{Carmi:2019cub} for its detailed form.


\section{Dispersion relation in $1$-$d$ CFTs}
\label{sec:deq1}

Consider a CFT in $d=1$ space-time dimensions, containing a primary operator $\phi$ with scaling dimension $\D_\phi$.
The goal of this section is to derive the dispersion relation for the four-point function $\langle \phi(x_1) \phi(x_2) \phi(x_3) \phi(x_4) \rangle $ of identical primary operators $\phi$. This four-point function is fixed by conformal symmetry, up to a function of a single cross-ratio $z$ as follows,
\begin{equation}
\label{4ptcorr}
\langle \phi(x_1) \phi(x_2) \phi(x_3) \phi(x_4) \rangle = \frac{\mm{G}(z)}{|x_{12}|^{2\Delta_{\phi}}|x_{34}|^{2\Delta_{\phi}}} \  , \quad \textrm{where, } \ \ z=\frac{x_{12}x_{34}}{x_{13}x_{24}} \ .
\end{equation}

\noindent Crossing symmetry of the correlator implies the following constraint on the function $\mathcal{G}(z)$, 
\begin{equation}
\label{crossingsym}
    \begin{split}
         z^{-2\Delta_{\phi}}\mathcal{G}(z)=(1-z)^{-2\Delta_{\phi}}\mathcal{G}(1-z) \quad \text{for} \ z\in(0,1) \ .
    \end{split}
\end{equation} 

\noindent The function $\mathcal{G}(z)$ can be expanded using the s-channel OPE (i.e. $x_2 \rightarrow x_1$) in terms of $SL(2,\mathbb{R})$ conformal blocks ($G_{\Delta} (z)$),
\begin{equation}
\label{cbexpsch}
\mathcal{G}(z) = \sum_{\Delta_{\mathcal{O}}} \left(c_{\phi\phi\mathcal{O}}\right)^{2} G_{\Delta_{\mathcal{O}}} (z)
\end{equation} 

\noindent where $c_{\phi\phi\mathcal{O}}$ are the OPE coefficients. The conformal partial wave expansion that will be useful for deriving the dispersion relation using the Lorentzian inversion formula (LIF) is \cite{Simmons-Duffin:2017nub,Mazac:2018qmi}:
\begin{align}
\label{eq:inv0}
\mathcal{G}(z) = \int_{\frac{1}{2}-i\infty}^{\frac{1}{2}+i\infty} \frac{d \D}{2\pi i} \frac{I_\D}{2 K_\D} G_\D(z) + \frac{1}{2\pi^2}\sum_{m \in 2N} \frac{\G^2(m)}{\G(2m-1)} \tilde{I}_m G_m(z) \quad \textrm{for } \  z \in (0,1) 
\end{align}

\noindent where the first term corresponds to the principal series representation of $SL(2,\mathbb{R})$, the second sum involves the discrete series representation of $SL(2,\mathbb{R})$, and $K_{\Delta}$ is defined as,
\begin{align}
K_{\Delta} = \frac{\sqrt{\pi} \G(\D-\frac{1}{2})\G^2(\frac{1-\D}{2})}{\G(1-\D)\G^2(\frac{\D}{2})} \ .
\end{align}

\noindent The coefficients $\left(c_{\phi\phi\mathcal{O}}\right)^{2}$ in the conformal block decomposition Eq.~\eqref{cbexpsch} are given by the residue of the simple pole of the function $\frac{I_\D}{2 K_\D}$ at $\Delta=\Delta_{O}$ where $\Delta_{O}$ is the scaling dimension of the operator $\mathcal{O}$ exchanged in the s-channel OPE. 

\vskip 4pt
\noindent The Lorentzian inversion formula (LIF) for the discrete series coefficient is \cite{Simmons-Duffin:2017nub,Mazac:2018qmi}:
\begin{align}
\tilde{I}_m &= \frac{4\G^2(m)}{\G(2m)} \int_0^1 dw w^{-2}G_m(w)\mathrm{dDisc}[\mm{G}(w)] \label{eq:inv1} 
\end{align}
where $G_m(w)$ is the conformal block with $m$ being an integer. The Lorentzian inversion formula (LIF) for the principal series OPE coefficient function is \cite{Mazac:2018qmi}:
\begin{align}
I_\D  &= 2 \int_0^1 dw w^{-2} H_{\D}(\Delta_{\phi}, w) \mathrm{dDisc}[\mm{G}(w)] \ .\label{eq:inv2}
\end{align}

\noindent In the LIF above, $\mathrm{dDisc}[\mm{G}(w)]$ is the double discontinuity of the 4-point function, defined for boson and fermions as:
\begin{align}
\label{dDiscBbf}
      &  \mathrm{dDisc}_{B}\left[\mathcal{G}(z)\right]  = \mathcal{G}(z) - \frac{\widetilde{\mathcal{G}}\left(\frac{1}{z}+i\epsilon\right)+ \widetilde{\mathcal{G}}\left(\frac{1}{z}-i\epsilon\right)}{2} \nonumber \\
      & \mathrm{dDisc}_{F}\left[\mathcal{G}(z)\right] = \mathcal{G}(z) + \frac{\widetilde{\mathcal{G}}\left(\frac{1}{z}+i\epsilon\right)+ \widetilde{\mathcal{G}}\left(\frac{1}{z}+i\epsilon\right)}{2}
\end{align}

\noindent where $z\in(0,1)$ and $\widetilde{\mathcal{G}}\left(z\right)=z^{-2\Delta_{\phi}}\mathcal{G}\left(z\right)$.

The inversion formula kernel $H_{\D}(\Delta_{\phi}, w)$ depends on whether the $\phi$'s are bosons or fermions. It also depends explicitly on the scaling dimension $\Delta_{\phi}$. We will subsequently use the notation $H^{B}_{\D}(\Delta_{\phi}, w)$ for bosons, and $H^{F}_{\D}(\Delta_{\phi}, w)$ for fermions. The inversion kernel was explicitly computed in \cite{Mazac:2018qmi} for bosons with scaling dimensions $\D_\phi$ being integers, and for fermions with scaling dimensions $\D_\phi$ being half-integers. As opposed to the inversion formula in higher $d$, the $1$-$d$ Lorentzian inversion formula manifests crossing symmetry \cite{Mazac:2018qmi}.

As shown in \cite{Carmi:2019cub} and reviewed in Section~\ref{sec:dg2}, for $d\geq 2$ one can derive a dispersion relation by combining the Lorentzian inversion formula and the conformal partial wave decomposition. We will now proceed to do an analogous computation for $1$-$d$ CFT. We start from the $1$-$d$ Lorentzian inversion formulae Eqs.~\eqref{eq:inv1} and \eqref{eq:inv2}, and insert it in the conformal block expansion Eq.~\eqref{eq:inv0}. Then exchanging the order of integrations gives the $1$-$d$ dispersion relation:
\begin{align}\label{Dispersion_Relation}
\boxed{ \mathcal{G}(z) = \int_0^1 dw  \ K(\Delta_{\phi},z,w) \ \mathrm{dDisc}[\mm{G}(w)]  } 
\end{align}

\noindent Here the dispersion kernel $K(\Delta_{\phi},z,w)$ consists of the following two components:
\begin{equation}
 K(\Delta_{\phi},z,w) =  K_{dis} (z,w) + K_{princ} (\Delta_{\phi}, z,w)
\end{equation}
with $K_{dis} (z,w)$ and $K_{princ} (\Delta_{\phi}, z,w)$ being the kernels corresponding to  the discrete and principal series respectively and are given as,
\begin{align}
K_{dis} (z,w) &=  \frac{2}{\pi^2}  w^{-2}  \sum_{m \in 2N} \frac{\G^4(m)}{\G(2m)\G(2m-1)} G_m(w) G_m(z)  \label{eq:t1} \\
K_{princ} (\Delta_{\phi},z,w) &=  \frac{1}{2\pi} w^{-2}  \int_{\frac{1}{2}-i\infty}^{\frac{1}{2}+i\infty} \frac{d \D}{2\pi i}    \frac{\G^2(\D)}{\G(2\D-1)} \frac{\cos[\frac{\pi \D}{2}]}{\sin[\frac{\pi \D}{2}]} H_{\D}(\Delta_{\phi}, w)   G_\D(z) \ . \label{eq:t2}
\end{align}

\noindent Note that the discrete series kernel does not depend on the function $H_{\D}(\Delta_{\phi}, w)$. This implies that $K_{dis}(z,w)$ is the same for bosons and fermions. On the other hand, the principal series kernel $K_{princ} (\Delta_{\phi},z,w)$ depends on  $H_{\D}(\Delta_{\phi}, w)$ which is different for bosons and fermions. Consequently, the full dispersion kernel $K(\Delta_{\phi},z,w)$ also depends on whether we consider bosons or fermions. The external scaling dimension $\Delta_{\phi}$ dependence of the dispersion kernel comes only from the principal series term. Let us also note that since the OPE coefficient function obtained by applying the LIF of \cite{Mazac:2018qmi} is a fully crossing symmetric object, the dispersion relation Eq.~\eqref{Dispersion_Relation} should manifest the crossing symmetry property Eq.~\eqref{crossingsym} of $\mathcal{G}(z)$. This implies that the kernel in the dispersion relation must satisfy the relation:
\begin{equation}
    \begin{split}
        z^{-2\Delta_{\phi}} K(\Delta_{\phi},z,w)=(1-z)^{-2\Delta_{\phi}} K(\Delta_{\phi},1-z,w) \ .
    \end{split}
\end{equation}
In the following subsections, we analytically compute the two kernels  $K_{dis} (z,w)$ and $K_{princ} (\D_{\phi}, z,w)$ in Eqs.~\eqref{eq:t1}-\eqref{eq:t2}. We will denote the bosonic and fermionic kernels as $K^{B}(\Delta_{\phi},z,w)$ and $K^{F}(\Delta_{\phi},z,w)$ respectively. We will see that writing the sums in terms of Legendre functions will facilitate computing them analytically.

\subsection{The discrete series kernel}\label{Discrete_Kernel}

The kernel corresponding to the discrete series is Eq.~\eqref{eq:t1}:
\begin{align}
\label{eq:cl1}
K_{dis} (z,w) &=  \frac{2}{\pi^2}  w^{-2}  \sum_{m \in 2N} \frac{\G^4(m)}{\G(2m)\G(2m-1)} G_m(w) G_m(z) 
\nn 
&=  \frac{8}{\pi^2} w^{-2}  \sum_{m \in 2N} (2m-1)Q_{m-1} (\hat w) Q_{m-1} (\hat z)
\end{align}
where, in the second line we wrote the conformal block in terms of a Legendre function of the second type (LegendreQ's) as follows: 
\begin{align}
\label{eq:q1f}
G_\D (w) = w^\D {}_2 F_1 (\D,\D,2\D,w) =  \frac{2\G(2\D)}{\G^2(\D)} Q_{\D-1}(\hat w)
\end{align}
with the hatted variables defined as $\hat{w} = \frac{2}{w} -1$ and  $\hat{z} = \frac{2}{z} -1$. We compute the sum over LegendreQ's in the second line of Eq.~\eqref{eq:cl1} in Eq.~\eqref{sumQQ}. The result of this sum is,
\begin{align}
K_{dis} (z,w)= \frac{2}{\pi^2} w^{-2}  \bigg( \frac{\log\Big[ \frac{\hat w+1}{\hat w-1} \frac{\hat z-1}{\hat z+1} \Big]}{\hat z-\hat w} + \frac{\log\Big[ \frac{-\hat w+1}{-\hat w-1} \frac{\hat z-1}{\hat z+1} \Big]}{\hat w+\hat z} \bigg) 
\end{align}
In terms of the cross ratios $w$ and $z$ we have,
\begin{align}
\label{Kdisc}
K_{dis} (z,w)= \frac{z}{\pi^2 w} \bigg[  \bigg( \frac{1}{  w-  z} + \frac{1}{  w +(1-w) z} \bigg) \log\big(1-z \big) 
\nn
+\bigg(- \frac{1}{  w-  z} + \frac{1}{  w +(1-w) z} \bigg) \log\big(1-w \big)\bigg]
\end{align} 

\noindent The discrete kernel is independent of the external scaling dimensions $\D_\phi$, as mentioned before.

\subsection{The principal series kernel: Bosons}
\label{subsec:KBprinc}

The bosonic kernel corresponding to the principal series Eq.~\eqref{eq:t2} is:
\begin{align}
\label{eq:pl1}
K^{B}_{princ} (\Delta_{\phi},z,w) &= \frac{1}{2\pi w^2}   \int_{\frac{1}{2}-i\infty}^{\frac{1}{2}+i\infty} \frac{d \D}{2\pi i}    \frac{\G^2(\D)}{\G(2\D-1)} \frac{\cos[\frac{\pi \D}{2}]}{\sin[\frac{\pi \D}{2}]} H^{B}_{\D}(\D_{\phi},w)   G_\D(z) 
\nn
&= \frac{1}{2\pi w^2}   \int_{\frac{1}{2}-i\infty}^{\frac{1}{2}+i\infty} \frac{d \D}{2\pi i}    \frac{\G^2(\D)}{\G(2\D-1)} \frac{\cos[\frac{\pi \D}{2}]}{\sin[\frac{\pi \D}{2}]} H^{B}_{\D}(\D_{\phi},w)  z^\D {}_2 F_1 (\D,\D,2\D,z) \ .
\end{align}
The function $H_{\Delta}^B(\D_{\phi},w)$ simplifies when the external scaling dimensions are integer numbers: $\D_\phi \in N$ and is computed in \cite{Mazac:2018qmi},
\begin{align}
\label{eq:pl2}
H_\D^B(\D_{\phi},w) &=  \frac{2\pi}{\sin(\pi \D)} \bigg[ w^{-2\D_{\phi}+2} p_\D(w) + \Big(\frac{w}{w-1}\Big)^{-2\D_{\phi}+2} p_\D \Big(\frac{w}{w-1} \Big) \bigg] \nonumber \\
&+ \frac{2\pi}{\sin(\pi \D)} q_\D^{\D_\phi}(w) 
\end{align}
where,
\begin{align}
p_\D(w) &= {}_2F_1(\D,1-\D,1,w) \notag \\ 
q_\D^{\D_\phi}(w) &= w^{-2\D_{\phi}+2} \Big[ a_\D^{\D_{\phi}}(w) + b_\D^{\D_{\phi}}(w) \log(1-w) \Big]
\end{align}
and $a_\D^{\Delta_{\phi}}(w)$ and $b_\D^{\Delta_{\phi}}(w)$ are polynomials in the variables $\D$ and $w$. Plugging Eq.~\eqref{eq:pl2} inside Eq.~\eqref{eq:pl1} gives:
\begin{align}
\label{eq:h5}
K^{B}_{princ}(\Delta_{\phi},z,w)  &= K^{(p)}_{princ}(\Delta_{\phi},z,w) + K_{princ}^{(q)}(\Delta_{\phi},z,w) \notag \\ 
&=  
\frac{1}{ 2w^2} \int\frac{d \D}{2\pi i} \bigg[   \frac{ w^{2-2\D_{\phi}}  \G^2(\D)}{\G(2\D-1)} \frac{{}_2F_1(\D,1-\D,1,w) }{\sin^2[\frac{\pi \D}{2}]}  z^\D {}_2 F_1 (\D,\D,2\D,z)  \nonumber \\ 
& \hspace{2.5cm}+ \Big(w \to \frac{w}{w-1} \Big) \bigg] \notag \\
& + \frac{1}{ 2w^2}  \int  \frac{d \D}{2\pi i}  \frac{    \G^2(\D)}{\G(2\D-1)} \frac{q_\D^{\D_\phi}(w)}{\sin^2[\frac{\pi \D}{2}]}  z^\D {}_2 F_1 (\D,\D,2\D,z) \ .
\end{align}

\noindent The kernels $K^{(p)}_{princ}(\Delta_{\phi},z,w)$ and $K^{(q)}_{princ}(\Delta_{\phi},z,w)$ can be written in terms of the Legendre functions of first kind (LegendreP's) and second kind (LengendreQ's) as,
\begin{align}
K^{(p)}_{princ}(\Delta_{\phi},z,w) &= 
w^{-2}\bigg[  w^{2-2\D_{\phi}}   \int_{\frac{1}{2}-i\infty}^{\frac{1}{2}+i\infty} \frac{d \D}{2\pi i}    \frac{1}{\sin^2[\frac{\pi \D}{2}]}  (2\D-1) P_{\D-1}(\tilde w) Q_{\D-1}(\hat z)  \nonumber \\
& \hspace{1.5cm}+\Big(w \to \frac{w}{w-1} \Big) \bigg] \label{eq:nb3} \\
K^{(q)}_{princ}(\Delta_{\phi},z,w) &= 
  w^{-2 }   \int_{\frac{1}{2}-i\infty}^{\frac{1}{2}+i\infty} \frac{d \D}{2\pi i}    \frac{1}{\sin^2[\frac{\pi \D}{2}]}  (2\D-1) q_\D^{\D_\phi}(w) Q_{\D-1}(\hat z) \label{eq:nb4}
\end{align}
where we used the following relations between the Legendre functions and hypergeometric functions:
\begin{align}
\label{eq:nb5}
 z^\D {}_2 F_1 (\D,\D,2\D,z) =  \frac{2\G(2\D)}{\G^2(\D)} Q_{\D-1}(\hat z) \  , \ \ \ \ \  {}_2F_1(\D,1-\D,1,w) =P_{\D-1}(\tilde w)
\end{align}
with, $\hat{z} = \frac{2}{z} -1$ and $\tilde{w} \equiv 1-2w$. We analytically compute Eqs.~\eqref{eq:nb3}, \eqref{eq:nb4} in appendix \ref{Dispersion_Kernel}. In particular, the kernel $K^{(p)}_{princ}(\D_{\phi}, z,w)$ is computed in Eq.~\eqref{Kpzw} and is given by,
\begin{align}
\label{eq:sdkbsd}
K^{(p)}_{princ}(\Delta_{\phi},z,w) & =  \frac{z}{\pi^2w }   \bigg[   w^{1-2\D_\phi}    \bigg(
\frac{\log \big(\frac{1-z}{z(1-w)}\big)}{w z-1}  - \frac{\log \big( z(1-w) \big)}{1+ z(w-1)}  \bigg) 
\nn
& +\Big(\frac{w}{w-1}\Big)^{1-2\D_{\phi}}    \bigg(
\frac{\log \big(\frac{(1-w)(1-z)}{z}\big)}{1-w(1-z)}  - \frac{\log \big(  \frac{1-w}{z} \big)}{1-w-z}  \bigg) \bigg] \ .
\end{align}

\noindent Closed form expressions for $K^{(q)}_{princ}(\Delta_{\phi},z,w)$ can be obtained on a case-by-case basis and are computed in appendix \ref{app:Kqprinc}. For example, for the case of $\Delta_\phi=1$ we have $a_\D^{\D_{\phi}}(w)=b_\D^{\D_{\phi}}(w)=0$ \cite{Mazac:2018qmi} and hence we get,
\begin{align}
K_{princ}^{(q)}(\Delta_{\phi}=1,z,w) = 0 \ , \qquad K^{B}_{princ}(\Delta_{\phi}=1,z,w) = K_{princ}^{(p)}(\Delta_{\phi}=1,z,w) \ .
\end{align}
The expressions for the kernel $K^{(q)}_{princ}(\Delta_{\phi},z,w)$, for $\Delta_{\phi}=2,3$ are given in equations Eq.~\eqref{Kqb2} and Eq.~\eqref{Kqb3} respectively.

\subsection{Improved bosonic kernel}
\label{subsec:KBimproved}

The kernel in the bosonic Lorentzian OPE inversion formula, which was used in the previous section to derive the dispersion relation, has the property:
\begin{equation}
\label{HBbound}
    \begin{split}
         H_{\Delta}^{B}(\D_{\phi},z)= O(z^{0}), \quad  \text{as} \ z\rightarrow 0.
    \end{split}
\end{equation}

\noindent As shown in \cite{Mazac:2018qmi}, a Lorentzian inversion formula involving a kernel that satisfies the above property can only be applied to correlators which obey the following super-boundedness condition in the Regge limit:
\begin{equation}
    \begin{split}
         |z^{-2\Delta_{\phi}} \mathcal{G}(z) | = O(|z|^{-1-\epsilon}), \quad  \text{as} \ z\rightarrow \infty, \text{with} \ \epsilon >0 .
    \end{split}
\end{equation}

\noindent Consequently, the dispersion kernel $K^{B}(\Delta_{\phi},z,w) $ derived in the previous section applies only to such super-bounded correlators. We will now derive a dispersion relation kernel that can be used for correlators which satisfy the weaker Regge bound:
\begin{equation}
    \begin{split}
         |z^{-2\Delta_{\phi}} \mathcal{G}(z) | = O(1), \quad  \text{as} \ z\rightarrow \infty.
    \end{split}
\end{equation}

\noindent In \cite{Mazac:2018qmi}, it was shown that an improved Lorentzian inversion formula applicable for Regge-bounded correlators can be obtained by constructing a new  inversion kernel $\widetilde{H}^{B}_{\Delta}(\D_{\phi},z)$ which is given by,
\begin{equation}
    \begin{split}
       \widetilde{H}^{B}_{\Delta}(\D_{\phi},z) = {H}^{B}_{\Delta}(\D_{\phi},,z) - \hat{H}^{corr}_{\Delta}(\D_{\phi},z)
           \end{split}
\end{equation}

\noindent where $\hat{H}^{corr}_{\Delta}(\D_{\phi},z)$ is a correction term, which is chosen so that in contrast to Eq.~\eqref{HBbound}, $\widetilde{H}^{B}_{\Delta}(\D_{\phi},z)$ behaves as,
\begin{equation}
\label{HBbound1}
    \begin{split}
        \widetilde{H}^{B}_{\Delta}(\D_{\phi},z)= O(z^{2}), \quad  \text{as} \ z\rightarrow 0.
    \end{split}
\end{equation}

\noindent In \cite{Mazac:2018qmi}, the correction term $\hat{H}^{corr}_{\Delta}(\D_{\phi},z)$  was chosen to be\footnote{This choice is not unique. $\frac{{H}^{B}_{\Delta}(z)}{\kappa_{\Delta}}$ has single poles and double poles at the double trace operators $\Delta^{B}_{n}=2\Delta_{\phi}+2n$ with residues $\hat{H}^{B}_{n,1}(z)$ and $\hat{H}^{B}_{n,2}(z)$ respectively. We can also choose the correction term to be proportional to $\hat{H}^{B}_{n,2}(z)$ with $n \neq 0$, or $\hat{H}^{B}_{n,1}(z)$ to modify the $z\rightarrow 0$ limit of the inversion kernel.} 
\begin{equation}
\label{HBcorr}
    \begin{split}
      \hat{H}^{corr}_{\Delta}(\D_{\phi},z) = \frac{\pi^{2}2^{2(\Delta_{\phi}-1)}\Gamma\left(\Delta_{\phi}+\frac{1}{2}\right)}{\Gamma\left(\Delta_{\phi}\right)^{3}\Gamma\left(2\Delta_{\phi}-\frac{1}{2}\right)} \frac{\Gamma\left(\Delta_{\phi}-\frac{\Delta}{2}\right)^{2}\Gamma\left(\Delta_{\phi}-\frac{1-\Delta}{2}\right)^{2}}{\Gamma\left(1-\frac{\Delta}{2}\right)^{2}\Gamma\left(\frac{1+\Delta}{2}\right)^{2}} \hat{H}^{B}_{0,2}(\D_{\phi},z)
    \end{split}
\end{equation}

\noindent where $\hat{H}^{B}_{0,2}(\D_{\phi},z)$ is 
\begin{equation}
\label{HB02}
    \begin{split}
       \hat{H}^{B}_{0,2}(\D_{\phi},z)& = \lim_{\Delta \to 2\Delta_{\phi}}\frac{d}{d\Delta} \left[(\Delta-2\Delta_{\phi})^{2}\frac{{H}^{B}_{\Delta}(\D_{\phi},z)}{\kappa_{\Delta}}\right] 
    \end{split}
\end{equation}

\noindent and $\kappa_{\Delta}=\frac{\sqrt{\pi}\Gamma\left(\Delta-\frac{1}{2}\right)\Gamma\left(\frac{1-\Delta}{2}\right)^{2}}{\Gamma(1-\Delta)\Gamma\left(\frac{\Delta}{2}\right)^{2}}$.  The improved inversion formula is then given by \cite{Mazac:2018qmi},
\begin{equation}
\label{bOPEinvimpr}
    \begin{split}
        I_{\Delta} & =2\int_{0}^{1} dw \ w^{-2} \widetilde{H}^{B}_{\Delta}(\D_{\phi},w) \mathrm{dDisc}_{B}[\mathcal{G}(w)] \\
        & + \lim_{\epsilon\to 0}\bigg[\int_{C_{\epsilon}^{+}} dw \ w^{-2} \widetilde{H}^{B}_{\Delta}(\D_{\phi},w) \mathcal{G}(w) + \int_{C_{\epsilon}^{-}} dw \ w^{-2} \widetilde{H}^{B}_{\Delta}(\D_{\phi},w) \mathcal{G}(w)\bigg]
    \end{split}
\end{equation}

\noindent where $C_{\epsilon}^{\pm}$ denote semicircular contours from $z=1+\epsilon$ to $z=1-\epsilon$ lying above and below the real axis respectively. Thus the LIF suitable for Regge-bounded correlators contains, in addition to the usual integral involving the double discontinuity, an additional contribution given by the contour integrals in the second line of the above formula.  

\noindent Using Eq.~\eqref{bOPEinvimpr} we then obtain the following dispersion relation for Regge-bounded correlators,
\begin{align}
\label{bbOPEinvimpr}
\mm{G}(z) & = \int_0^1 dw   \ \widetilde{K}^{B}(\Delta_{\phi}, z,w) \mathrm{dDisc_{B}}[\mm{G}(w)] \nonumber  \\
& +  \lim_{\epsilon\to 0}\bigg[\int_{C_{\epsilon}^{+}} dw \ \widetilde{K}^{B}(\Delta_{\phi}, z,w)\mathcal{G}(w) + \int_{C_{\epsilon}^{-}} dw \ \widetilde{K}^{B}(\Delta_{\phi}, z,w) \mathcal{G}(w)\bigg].
\end{align}

\noindent Here $\widetilde{K}^{B}(\Delta_{\phi}, z,w)$ is the improved dispersion kernel which is given by,
\begin{equation}
\label{KBimp}
    \begin{split}
\widetilde{K}^{B}(\Delta_{\phi}, z,w) = K^{B}(\Delta_{\phi}, z,w) -\hat{K}^{B}_{corr}(\Delta_{\phi}, z,w) 
 \end{split}
\end{equation}

\noindent and $\hat{K}^{B}_{corr}(\Delta_{\phi}, z,w)$ is the correction term:
\begin{equation}
\label{KBcorr}
    \begin{split}
\hat{K}^{B}_{corr}(\Delta_{\phi}, z,w)= \frac{1}{2\pi w^2}   \int_{\frac{1}{2}-i\infty}^{\frac{1}{2}+i\infty} \frac{d \D}{2\pi i}    \frac{\G^2(\D)}{\G(2\D-1)} \frac{\cos[\frac{\pi \D}{2}]}{\sin[\frac{\pi \D}{2}]} \hat{H}^{corr}_{\Delta}(\D_{\phi},w)  G_\D(z) .
 \end{split}
\end{equation}

\noindent As an example, we will now compute $\hat{K}^{B}_{corr}(\Delta_{\phi}, z,w)$ for $\Delta_{\phi}=1$. This result will be subsequently useful for testing the dispersion relation in Section \ref{sec:checks}.  In this case, we have from Eq.~\eqref{HB02}, 
\begin{equation}
\label{HBcorrdphi2}
    \begin{split}
      \hat{H}^{B}_{0,2}(\Delta_{\phi}=1,z) & = \lim_{\Delta \to 2}\frac{d}{d\Delta} \left[(\Delta-2)^{2}\frac{{H}^{B}_{\Delta}(\Delta_{\phi}=1,z)}{\kappa_{\Delta}}\right] \\
      & = \frac{2 (z^{2}- z+1)}{\pi ^2  (1-z)} \ .
    \end{split}
\end{equation}

\noindent Using the above result in Eq.~\eqref{HBcorr} gives, 
\begin{equation}
\label{HBcorrdphi1}
    \begin{split}
      \hat{H}^{corr}_{\Delta}(\Delta_{\phi}=1, z) & = \frac{2\pi^{3}}{\sin(\pi \Delta)}  \hat{H}^{B}_{0,2}(\Delta_{\phi}=1,z)\\
      & = \frac{4\pi }{\sin(\pi \Delta)}\frac{(z^{2}- z+1)}{ (1-z)} \ .
    \end{split}
\end{equation}

\noindent Then using Eq.~\eqref{HBcorrdphi1}, the correction term in the dispersion kernel for $\Delta_{\phi}=1$ is given by,  
\begin{equation}
\label{KBcorrdphi1}
    \begin{split}
 \hat{K}^{B}_{corr}(\Delta_{\phi}=1, z,w) & = \frac{1}{2\pi w^2}   \int_{\frac{1}{2}-i\infty}^{\frac{1}{2}+i\infty} \frac{d \D}{2\pi i}    \frac{\G^2(\D)}{\G(2\D-1)} \frac{\cos[\frac{\pi \D}{2}]}{\sin[\frac{\pi \D}{2}]} \hat{H}^{corr}_{\Delta}(\Delta_{\phi}=1,w)  G_\D(z) \\
& = - \frac{2 z(w^{2}- w+1)}{ \pi^{2}(1-w)w^{2}} \left[\log\left(\frac{1-z}{z}\right)+\frac{\log(z)}{1-z}\right] \ .
 \end{split}
\end{equation}

\noindent Let us also note that the above result can be expressed as,
\begin{equation}
\label{KBcorrdphi2}
    \begin{split}
 \hat{K}^{B}_{corr}(\Delta_{\phi}=1, z,w) & = - \frac{z^{2}(w^{2}- w+1)}{ \pi^{2}(1-w)w^{2}} \ \bar{D}_{1111}(z)
 \end{split}
\end{equation}

\noindent where
\begin{equation}
\label{Dbardphi1}
    \begin{split}
 \bar{D}_{1111}(z) = \frac{2 \log(z)}{z-1} -\frac{2\log(1-z)}{z}
 \end{split}
\end{equation}

\noindent is the reduced $D$-function. The latter is the non-derivative $4$-point contact Witten diagram in $AdS_{2}$ for massless scalars corresponding to a dual CFT$_{1}$ operator of dimension $\Delta_{\phi}=1$. 


\subsection{The principal series kernel: Fermions}
\label{subsec:KFprinc}

We now consider the computation of the principal series kernel for the case of fermions. Here the LIF kernel is given by \cite{Mazac:2018qmi},
\begin{align}
H_\D^F(\Delta_{\phi}, w)= &- \frac{2\pi}{\sin(\pi \D)} \bigg[ w^{-2\D_{\phi}+2} p_\D(w) + \Big(\frac{w}{w-1}\Big)^{-2\D_{\phi}+2} p_\D \Big(\frac{w}{w-1} \Big) \bigg] \nonumber \\
& - \frac{2\pi}{\sin(\pi \D)} \widehat{q}_\D^{\D_\phi}(w) \ .
\end{align}

\noindent Notice that the term in the square brackets above is the same as in the bosonic case of Eq.~\eqref{eq:pl2}, and $\widehat{q}_\D^{\D_\phi}(w) $ is:
\begin{align}
 \widehat{q}_\D^{\D_\phi}(w) = w^{-2\D_{\phi}+2} \Big[ \widehat{a}_\D^{\D_{\phi}}(w) + \widehat{b}_\D^{\D_{\phi}}(w) \log(1-w) \Big]
\end{align}
where $\widehat{a}_\D^{\D_{\phi}}$ and $\widehat{b}_\D^{\D_{\phi}}$ are polynomials in the variables $\D$ and $w$, which are different from the bosonic case. This implies the principal series kernel for fermions is: 
\begin{align}
\label{eq:h5}
&K^{F}_{princ}(\Delta_{\phi},z,w)  =   - K^{(p)}_{princ}(\Delta_{\phi},z,w) - \widehat{K}_{princ}^{(q)}(\Delta_{\phi},z,w) 
\end{align}
where, $K^{(p)}_{princ}(\Delta_{\phi},z,w)$ is the same as in Eq.~\eqref{eq:sdkbsd}, and $ \widehat{K}_{princ}^{(q)}(\Delta_{\phi},z,w) $ is defined by Eq.~\eqref{eq:nb4} with the replacement $q_\D^{\D_\phi}(w)  \rightarrow \widehat{q}_\D^{\D_\phi}(w) $. For the case of $\Delta_\phi=\frac{1}{2}$ we have $\widehat{a}_\D^{\D_{\phi}}=\widehat{b}_\D^{\D_{\phi}}=0$ \cite{Mazac:2018qmi}, and hence we have,
\begin{align}
\label{Kfp}
\widehat{K}_{princ}^{(q)} \left(\Delta_{\phi}=\frac{1}{2},z,w \right) = 0 \ ,  K^{F}_{princ} \left(\Delta_{\phi}=\frac{1}{2},z,w \right) = -K_{princ}^{(p)} \left(\Delta_{\phi}=\frac{1}{2},z,w \right) \ .
\end{align}

\noindent The expressions for the kernel $\widehat{K}^{(q)}_{princ}(\Delta_{\phi},z,w)$ for $\Delta_{\phi}=3/2, 5/2$ are given in Eqs.~\eqref{Kqf3by2} and \eqref{Kqf5by2} respectively.

\subsection{Dispersion relation from Polyakov bootstrap}
\label{sec:Pauloscomp}

In this section, we briefly review the approach, based on the Polyakov bootstrap, taken in \cite{Paulos:2020zxx} for deriving dispersion relations for $1$-$d$ CFTs. Our purpose here is to highlight the difference between this method and that of ours which instead employs the Lorentzian inversion formula.

\vskip 4pt
The Polyakov bootstrap in the case of $1d$ CFTs is the statement that the expansion of a four-point CFT correlator $ \hat{\mathcal{G}}(z) =\mathcal{G}(z)/z^{2\Delta_{\phi}}$ in terms of conformal blocks $G_{\Delta}(z)$ is equivalent to expanding the correlator in terms of Polyakov blocks $\mathcal{P}_{ \Delta_{O}}(\Delta_{\phi},z)$, which are crossing symmetric sums of exchange Witten diagrams in $AdS_{2}$. 
\begin{align}
\label{Polyakovboots}
    &   \hat{\mathcal{G}}(z) = \sum_{ \Delta_{O}} a_{ \Delta_{O}} \frac{G_{\Delta}(z)}{z^{2\Delta_{\phi}}} = \sum_{ \Delta_{O}} a_{ \Delta_{O}} \mathcal{P}_{ \Delta_{O}}(\Delta_{\phi},z) .
\end{align}

\noindent Here $a_{ \Delta_{O}} $ denote squared OPE coefficients. In \cite{Paulos:2020zxx}, it was shown that Polyakov blocks admit the following integral representations in terms of the double-discontinuity of the conformal blocks,
\begin{align}
\label{PBdisp}
    &   \mathcal{P}_{B, \Delta_{O}}(\Delta_{\phi},z) = - \int_{0}^{1} dw \ \widehat{g}_{B}(\Delta_{\phi},z,w) \ \mathrm{dDisc}_{B}\left[  \frac{G_{\Delta}(w)}{w^{2\Delta_{\phi}}}\right], \quad \text{for} \ \Delta_{O}>2\Delta_{\phi}, \\
    \label{PFdisp}
       &  \mathcal{P}_{F, \Delta_{O}}(\Delta_{\phi},z) =  \int_{0}^{1} dw \ \widehat{g}_{F}(\Delta_{\phi},z,w) \ \mathrm{dDisc}_{F}\left[ \frac{G_{\Delta}(w)}{w^{2\Delta_{\phi}}}\right], \quad \text{for} \ \Delta_{O}>2\Delta_{\phi}-1,
\end{align}

\noindent where the subscripts $B, F$ on the Polyakov blocks refer to the bosonic and fermionic cases respectively and $\widehat{g}_{B}(\Delta_{\phi},z,w), \ \widehat{g}_{F}(\Delta_{\phi},z,w)$ are certain kernels. The above integral representations were obtained in \cite{Paulos:2020zxx} by defining the Polyakov blocks in terms of a set of master functionals. These are extremal functionals which yield rigorous upper and lower bounds on the values of the four-point correlator $ \mathcal{G}(z)$. The action of such master functionals on conformal blocks $G_{\Delta}(z)$ can be expressed via contour integrals involving the kernels $\widehat{g}_{B}(\Delta_{\phi},z,w), \ \widehat{g}_{F}(\Delta_{\phi},z,w)$ in the bosonic and fermionic cases respectively. Then, one can arrive at Eqs.~\eqref{PBdisp}, \eqref{PFdisp} via an appropriate contour deformation. We refer the reader to \cite{Paulos:2020zxx} for the details of this derivation. In particular, they showed that these kernels satisfy following integral equation.
\begin{align}\label{gwidehatPaulos}
\widehat{g}(\Delta_{\phi},z,w) &= R_{\Delta_{\phi}}(z,w) - \int_{0}^{1} dz' R_{\Delta_{\phi}}(z,z') \ \widehat{g}(\Delta_{\phi},z',w) 
\end{align}
with,
\begin{align}
R_{\Delta_{\phi}}(z,w) &= \frac{1}{\pi} \sqrt{\frac{z(1-z)}{w(1-w)}} \frac{1+w}{\left(z - \frac{w}{w-1} \right) \left( \frac{1}{1-w}-z\right)} (1-w)^{2\Delta_{\phi}-\frac{3}{2}}
\end{align}

\vskip 4pt
\noindent Now, inserting the above integral expressions for the Polyakov blocks in the Polyakov block expansion of $\mathcal{G}(z) $ in Eq.~\eqref{Polyakovboots} gives the dispersion relations,
\begin{align}
\label{paulosdisp}
    &    \overline{\mathcal{G}}(z) = - \int_{0}^{1} dw \ \widehat{g}_{B}(\Delta_{\phi},z,w) \ \mathrm{dDisc}_{B}\left[ \overline{\mathcal{G}}(w)\right],  \\
       &    \widetilde{\mathcal{G}}(z) = \int_{0}^{1} dw \ \widehat{g}_{F}(\Delta_{\phi},z,w) \ \mathrm{dDisc}_{F}\left[ \widetilde{\mathcal{G}}(w)\right], 
\end{align}

\noindent where we have defined,
\begin{align}
\label{paulosdisp}
    &   \overline{\mathcal{G}}(z) =  \hat{\mathcal{G}}(z)-\sum_{ 0\le \Delta_{O}\le 2\Delta_{\phi}} a_{ \Delta_{O}} \mathcal{P}_{B, \Delta_{O}}(\Delta_{\phi},z) ,  \\
       &   \widetilde{\mathcal{G}}(z) =  \hat{\mathcal{G}}(z)-\sum_{ 0\le \Delta_{O}\le 2\Delta_{\phi}-1} a_{ \Delta_{O}} \mathcal{P}_{F, \Delta_{O}}(\Delta_{\phi},z) . 
\end{align}

\noindent We expect the kernels in the above dispersion relations to be identical to the kernels we have obtained in the previous section using the $1$-$d$ Lorentzian OPE inversion formula. In particular, we have checked that the kernel in the fermionic case for $\Delta_{\phi}=1/2$ obtained in this paper is the same as the corresponding expression found in \cite{Paulos:2020zxx}\footnote{The dispersion kernel defined in this paper is related to the one defined in \cite{Paulos:2020zxx} upto a pre-factor. This is because the cross-ratio dependent function in the correlator in \cite{Paulos:2020zxx} denoted as  $\hat{\mathcal{G}}(z)$ therein is related to the one we defined in Eq.~\eqref{4ptcorr} (i.e. $\mathcal{G}(z)$), as $\hat{\mathcal{G}}(z)=\mathcal{G}(z)/z^{2\Delta_{\phi}}$. We have taken this into account while making the comparisons.}.

\section{Checks for mean field theory correlators}
\label{sec:checks}

In this section, we perform checks of the dispersion relations derived in the previous sections. We choose our test correlators to be those of generalised free field theory (GFF). Since the bosonic (fermionic) double discontinuity annihilates the bosonic (fermionic) double trace conformal blocks, we will test the bosonic dispersion relation using the fermionic GFF correlator and check the fermionic dispersion relation using the bosonic GFF correlator. 

\subsection{Check for $\Delta_{\phi}=1$}
\label{gffcheck}

Consider the $1$-$d$ fermionic GFF correlator with $\Delta_{\phi}=1$ which is given by,
\begin{equation}
\label{gf}
    \begin{split}
        \mathcal{G}_{F}(z) =1- z^{2}+\left(\frac{z}{1-z}\right)^{2}.
    \end{split}
\end{equation}

\noindent The correlator $ \mathcal{G}_{F}(z)$ is not normalizable with respect to the inner product defined on the space of $SL(2,\mathbb{R})$ invariant functions in \cite{Mazac:2018qmi}, which is used to extract the OPE coefficient function from the conformal partial wave expansion. Therefore, the LIF and consequently the dispersion relation cannot be applied unless we consider a subtracted correlator where the non-normalizable contributions have been removed. Following \cite{Mazac:2018qmi,Paulos:2020zxx}, we then define a new correlator $\overline{\mathcal{G}}_{F}(z)$ where we take the subtraction term to be the crossing symmetric Polyakov blocks for the non-normalizable operators. $\overline{\mathcal{G}}_{F}(z)$ is also crossing symmetric by construction and is given by,
\begin{equation}
\label{gfsub}
    \begin{split}
        \overline{\mathcal{G}}_{F}(z) = \mathcal{G}_{F}(z) - \sum_{0\le \D_{\mathcal{O}} \le 2\Delta_{\phi}} \left(c_{\phi\phi\mathcal{O}}\right)^{2} \mathcal{P}_{B,\D_{\mathcal{O}}}(\Delta_{\phi}, z)
    \end{split}
\end{equation}

\noindent where $\mathcal{P}_{B,\D_{\mathcal{O}}}(\Delta_{\phi}, z)$ is the crossing symmetric bosonic Polyakov block for exchange scaling dimension $\Delta_{\mathcal{O}}$ and $c_{\phi\phi\mathcal{O}}$ denotes the OPE coefficients appearing in the spectrum of $\mathcal{G}_{F}(z)$. For the fermionic GFF correlator with $\Delta_{\phi}=1$, only the identity operator appears in the range $[0,2\Delta_{\phi}]$. Therefore in this case, only the identity bosonic Polyakov block $\mathcal{P}_{B,\D_{\mathcal{O}}=0}(\Delta_{\phi}=1,z)$ contributes to the sum in Eq.~\eqref{gfsub}. Now $\mathcal{P}_{B,\D_{\mathcal{O}}=0}(\Delta_{\phi}=1,z)$  is equal to the bosonic GFF correlator for $\Delta_{\phi}=1$ \cite{Mazac:2018qmi,Paulos:2020zxx},
\begin{equation}
\label{pblockb}
    \begin{split}
       \mathcal{P}_{B,\D_{\mathcal{O}}=0}(\Delta_{\phi}=1, z) =  1 +z^{2}+\left(\frac{z}{1-z}\right)^{2}.
    \end{split}
\end{equation}

\noindent Thus we have,
\begin{equation}
\label{gfsub1}
    \begin{split}
        \overline{\mathcal{G}}_{F}(z) &= \mathcal{G}_{F}(z) - \mathcal{P}_{B,\D_{\mathcal{O}}=0}(\Delta_{\phi}=1, z)= -2 z^{2}. 
    \end{split}
\end{equation}

\noindent where we have taken $c_{\phi\phi\mathbf{1}}=1$. To apply the dispersion relation, we need the bosonic double discontinuity of $\overline{\mathcal{G}}_{F}(z)$. This is given by,
\begin{equation}
\label{dDiscBGf}
    \begin{split}
        \mathrm{dDisc}_{B}\left[\overline{\mathcal{G}}_{F}(z)\right] & = \overline{\mathcal{G}}_{F}(z) - \frac{\overline{\mathcal{G}}^{+}_{F}(z+i\epsilon)+\overline{\mathcal{G}}^{+}_{F}(z-i\epsilon)}{2}
    \end{split}
\end{equation}

\noindent where $\overline{\mathcal{G}}^{+}_{F}(z) = z^{2\Delta_{\phi}}\overline{\mathcal{G}}_{F}\left(\frac{1}{z}\right)$ for $z\in (1,\infty)$. Then using Eq.~\eqref{gfsub1} we get\footnote{The bosonic double discontinuity of $\overline{\mathcal{G}}_{F}$ is the same as that of $\mathcal{G}_{F}$ since $\mathrm{dDisc}_{B}\left[\mathcal{P}_{B,\D}(\Delta_{\phi}, z)\right] =0$.},
\begin{equation}
\label{dDiscBGf1}
    \begin{split}
        \mathrm{dDisc}_{B}\left[\overline{\mathcal{G}}_{F}(z)\right] & =2(1-z^{2}).
    \end{split}
\end{equation}


\noindent Since $ \overline{\mathcal{G}}_{F}(z)$ is bounded but not super-bounded in the Regge limit, we will use the improved dispersion relation derived in Section \ref{subsec:KBimproved}.  The kernel appearing in this dispersion relation is
\begin{equation}
\label{Kbimpdphi1}
    \begin{split}
\widetilde{K}^{B}(\Delta_{\phi}, z,w)  & = K^{B}(\Delta_{\phi},z,w) -\hat{K}^{B}_{corr}(\Delta_{\phi} ,z,w) \ .
 \end{split}
\end{equation}

\noindent For $\Delta_{\phi}=1$,  using Eq.~\eqref{Kdisc} and Eq.~\eqref{eq:sdkbsd} we have,
\begin{align}
\label{KbDphi1}
& K^{B}(\Delta_{\phi}=1, z,w)  = \frac{z^2\left(w^2-w+1\right)  \left(w^2 z^2-2 w^2 z+w^2-2 w z+2 z\right)\log(1-z)}{\pi ^2 w^2 (w-z) (w z-1) (w z-w+1) (w z-w-z)}\nonumber \\
& \hspace{0.7cm}- \frac{z^2\left(w^2-w+1\right)  \left(w^2 z^2+2 w z-2 w-2 z+2\right)\log(z)}{\pi ^2 w^2 (w+z-1) (w z-1) (w z-w+1) (w z-z+1)} \nonumber \\
& \hspace{0.7cm} -\frac{(w-2) (w-1) (w+1) (2 w-1) z^2 \left(z^2-z+1\right)^2\log(1-w)}{\pi ^2 w (w-z) (w+z-1) (w z-1) (w z-w+1) (w z-z+1) (w z-w-z)}
\end{align}

\noindent and from Eq.~\eqref{KBcorrdphi1} we have,
\begin{equation}
\label{KcorrDphi1}
    \begin{split}
\hat{K}^{B}_{corr}(\Delta_{\phi}=1, z,w) & = -\frac{2 z(w^{2}- w+1)}{ \pi^{2}(1-w)w^{2}} \left[\log\left(1-z\right)+\frac{z \log(z)}{1-z}\right].
 \end{split}
\end{equation}

\noindent Now using the above expressions for the kernel and the double discontinuity we have checked that the following relation is satisfied,
\begin{equation}
    \begin{split}
\int_0^1 dw   \ \widetilde{K}^{B}(\Delta_{\phi}=1, z,w) \mathrm{dDisc_{B}}[\overline{\mm{G}}_{F}(w)] = -2 z^{2} = \overline{\mm{G}}_{F}(z).
 \end{split}
\end{equation}

\noindent This implies that for the test correlator under consideration, the terms in the dispersion relation Eq.~\eqref{bbOPEinvimpr} involving integrals over the contours $C^{\pm}_{\epsilon}$ must vanish in the limit $\epsilon \rightarrow 0$. We will now show that this is indeed the case.

\vskip 4pt
\noindent Let us define $w-1=\epsilon e^{i\phi}$. Then the integrals over the contours $C^{\pm}_{\epsilon}$ can be written as
\begin{align}
\label{cint}
& \lim_{\epsilon\to 0} i \epsilon \bigg[\int_{0}^{\pi} d\phi  e^{i\phi} \ \widetilde{K}^{B}(\Delta_{\phi}, z, \epsilon e^{i\phi}+1 ) \overline{\mathcal{G}}_{F}(\epsilon e^{i\phi}+1) \nonumber \\
& \hspace{2.0cm}+ \int_{2\pi}^{\pi}d\phi   e^{i\phi} \  \widetilde{K}^{B}(\Delta_{\phi}, z,\epsilon e^{i\phi}+1) \overline{\mathcal{G}}_{F}(\epsilon e^{i\phi}+1)\bigg]
\end{align}

\noindent To evaluate the above integral we only need to keep terms in the integrand which are singular as $\epsilon \rightarrow 0$. Now $ \overline{\mathcal{G}}_{F}(w) =-2w^{2}$ is finite for $w \rightarrow 1$. In the improved dispersion kernel,  the term $K^{B}(\Delta_{\phi}=1, z,w) $ as given in Eq.~\eqref{KbDphi1} is also regular for $w\rightarrow 1$.  But from Eq.~\eqref{KcorrDphi1} we see that the correction term $\hat{K}^{B}_{corr}(\Delta_{\phi}=1, z,w)$ has a simple pole at $w =1$.  Thus, only the $\hat{K}^{B}_{corr}(\Delta_{\phi}=1, z,w)$ term in the improved dispersion kernel can contribute to the integral in equation Eq.~\eqref{cint} as $\epsilon \rightarrow 0$. Keeping terms of the relevant order we then get,
\begin{align}
\label{cint1}
&  \lim_{\epsilon\to 0} i \epsilon \bigg[\int_{0}^{\pi} d\phi  e^{i\phi} \ \widetilde{K}_{B}(\Delta_{\phi}=1, z, \epsilon e^{i\phi}+1 ) \overline{\mathcal{G}}_{F}(\epsilon e^{i\phi}+1) \nonumber \\
& \hspace{1.2cm}+ \int_{2\pi}^{\pi}d\phi   e^{i\phi} \  \widetilde{K}_{B}(\Delta_{\phi}=1, z,\epsilon e^{i\phi}+1) \overline{\mathcal{G}}_{F}(\epsilon e^{i\phi}+1)\bigg] \nonumber \\
& =  \frac{4 z}{ \pi^{2}} \left[\log\left(1-z\right)+\frac{z \log(z)}{1-z}\right]  \bigg[ \int_{0}^{\pi} d\phi  + \int_{2\pi}^{\pi} d\phi \bigg]\nonumber \\
& =0.
\end{align}

\noindent Thus only the term involving the double discontinuity in the dispersion relation Eq.~\eqref{bbOPEinvimpr} is relevant for checking the bosonic dispersion relation for the test correlator in Eq.~\eqref{gfsub1}. 

\subsection{Check for $\Delta_{\phi}=\frac{1}{2}$}
\label{gfbcheck}

We will now check the fermionic dispersion relation for $\Delta_{\phi}=\frac{1}{2}$. As in section \ref{gfbcheck}, we take the test correlator to include subtractions to make it normalizable. Specifically, our test correlator is:
\begin{equation}
\label{gtb}
    \begin{split}
        \widetilde{\mathcal{G}}_{B}(z) =  \mathcal{G}_{B}(z)  - \sum_{0\le \Delta_{\mathcal{O}}\le 2\Delta_{\phi}-1} \left(c_{\phi\phi\mathcal{O}}\right)^{2} \mathcal{P}_{F,\D_{\mathcal{O}}}(\Delta_{\phi}, z) 
    \end{split}
\end{equation}

\noindent where $\mathcal{G}_{B}(z)$ is the $1$-$d$ CFT correlator for generalised free bosons with $\Delta_{\phi}=1/2$ which is given by,
\begin{equation}
\label{gbdphihalf}
    \begin{split}
        \mathcal{G}_{B}(z) =1 + z+\left(\frac{z}{1-z}\right)
    \end{split}
\end{equation}

\noindent and $\mathcal{P}_{F,\D_{\mathcal{O}}}(\Delta_{\phi}, z) $ is the crossing symmetric fermionic Polyakov block for exchange dimension $\Delta_{\mathcal{O}}$ and $c_{\phi\phi\mathcal{O}}$'s are the OPE coefficients in the spectrum of $\mathcal{G}_{B}(z)$. Now as in the previous case, for the bosonic GFF correlator with $\Delta_{\phi}=1/2$ only the identity operator is present in the range $[0,2\Delta_{\phi}-1]$. Consequently, in the subtraction term in Eq.~\eqref{gtb} we only have the identity fermionic Polyakov block $\mathcal{P}_{F,\D_{\mathcal{O}}=0}(\Delta_{\phi}=1/2,z)$ which is given by the fermionic GFF correlator for $\Delta_{\phi}=1/2$ \cite{Mazac:2018qmi,Paulos:2020zxx},
\begin{equation}
\label{Pf0}
    \begin{split}
         \mathcal{P}_{F,\Delta_{\mathcal{O}}=0}(\Delta_{\phi},z) = 1- z+ \left(\frac{z}{1-z}\right).
    \end{split}
\end{equation}

\noindent Then using Eqs.~\eqref{gbdphihalf} and \eqref{Pf0} we have,
\begin{equation}
\label{gtb1}
    \begin{split}
         \widetilde{\mathcal{G}}_{B}(z) =  \mathcal{G}_{B}(z)  - \mathcal{P}_{F,\Delta_{\mathcal{O}}=0}\left(\Delta_{\phi}=\frac{1}{2},z\right) = 2z.
    \end{split}
\end{equation}

\noindent Now to compute the fermionic double discontinuity of $ \widetilde{\mathcal{G}}_{B}(z)$ we use the definition:
\begin{equation}
\label{dDiscFGb}
    \begin{split}
        \mathrm{dDisc}_{F}\left[ \widetilde{\mathcal{G}}_{B}(z)\right] & =   \widetilde{\mathcal{G}}_{B}(z) - \frac{  \widetilde{\mathcal{G}}^{+}_{B}(z+i\epsilon)+  \widetilde{\mathcal{G}}^{+}_{B}(z-i\epsilon)}{2}
    \end{split}
\end{equation}

\noindent where $  \widetilde{\mathcal{G}}^{+}_{B}(z) = -z^{2\Delta_{\phi}}  \widetilde{\mathcal{G}}_{B}\left(\frac{1}{z}\right)$ for $z\in (1,\infty)$. For $   \widetilde{\mathcal{G}}_{B}(z)$ given by Eq.~\eqref{gtb1} we then obtain,
\begin{equation}
\label{dDiscFGb1}
    \begin{split}
        \mathrm{dDisc}_{F}\left[  \widetilde{\mathcal{G}}_{B}(z)\right] & =2(1+z) \ .
    \end{split}
\end{equation}

\noindent The dispersion kernel $K^{F}\left(\Delta_{\phi}=\frac{1}{2}, z,w\right)$ using Eqs.~\eqref{Kdisc}, \eqref{eq:sdkbsd}, \eqref{Kfp} is given by,
\begin{align}
\label{KfDphihalf}
& K^{F}\left(\Delta_{\phi}=\frac{1}{2}, z,w\right) = \frac{z^2 \left(w^2 z^2+2 w^2 z-2 w^2+2 w z^2-5 w z+5 w-2 z^2+5 z-5\right) \log (z)}{\pi ^2 (w+z-1) (w z-1) (w z-w+1) (w z-z+1)}\nonumber\\
& \hspace{0.7cm} +  \frac{(z-1) z \left(w^2 z^2-4 w^2 z+w^2+2 w z^2+w z+2 w-2 z^2-z-2\right) \log (1-z)}{\pi ^2 (w-z) (w z-1) (w z-w+1) (w z-w-z)}\nonumber \\
& \hspace{0.7cm} -\frac{(w-2) (w+1) (2 w-1) \left(w^2-w+1\right) (z-1) z^2 \left(z^2-z+1\right) \log (1-w)}{\pi ^2 w (w-z) (w+z-1) (w z-1) (w z-w+1) (w z-z+1) (w z-w-z)} \ .
\end{align}

\noindent With the above expressions for the kernel and $ \mathrm{dDisc}_{F}$ we have checked that,
\begin{equation}
    \begin{split}
\int_0^1 dw   \ K_{F}\left(\Delta_{\phi}=\frac{1}{2}, z,w\right) \mathrm{dDisc_{F}}[  \widetilde{\mathcal{G}}_{B}(w)] = 2 z =   \widetilde{\mathcal{G}}_{B}(z) \ .
 \end{split}
\end{equation}

\noindent This confirms that our test correlator indeed satisfies the $1$-$d$ fermionic dispersion relation. 


\section{Relation between the LIF kernel and the dispersion kernel}
\label{sec:HandKrel}

In this section, we obtain a relation between the inversion formula kernel $H_{\Delta}(\Delta_{\phi},w)$ and the dispersion relation kernel $K(\Delta_{\phi},z,w)$. We start with the Euclidean OPE inversion formula for the OPE coefficient function $I_{\Delta}$ which is given by,
\begin{align}
\label{euclOPEinv}
I_{\Delta} = \int_{-\infty}^{\infty} dz z^{-2} \ \psi_{\Delta}(z) \mathcal{G}(z) \ ,
\end{align}

\noindent where $\mathcal{G}(z)$ is the $1$-$d$ CFT correlator defined in Eq.~\eqref{4ptcorr}. $\psi_{\Delta}(z)$ is the conformal partial wave, which is defined for $z\in (0,1)$ as the following linear combination of the conformal block $G_{\D}(z)$ and the shadow conformal block $G_{1-\D}(z)$,
\begin{align}
\label{cpw}
& \psi^{(0)}_{\Delta}(z) =  K_{1-\D} G_{\D}(z)+K_{\D} G_{1-\D}(z) \quad \text{for} \ z\in (0,1) \ ,
\end{align}

\noindent where
\begin{align}
&  K_{\D} =\frac{\sqrt{\pi}\Gamma\left(\Delta-\frac{1}{2}\right)\Gamma\left(\frac{1-\Delta}{2}\right)^{2}}{\Gamma(1-\Delta)\Gamma\left(\frac{\Delta}{2}\right)^{2}} \ . 
\end{align}

\noindent For $z\in (-\infty,0)$ and $z\in (1,\infty)$ the conformal partial waves are given by \cite{Mazac:2018qmi},
\begin{align}
\label{relscpw}
& \psi^{(-)}_{\Delta}(z) =  \psi^{(0)}_{\Delta}\left(\frac{z}{z-1}\right) \quad \text{for} \ z\in (-\infty,0) \ , \nonumber \\
& \psi^{(+)}_{\Delta}(z) = \frac{1}{2}\left[\psi^{(0)}_{\Delta}(z+ i \epsilon)+ \psi^{(0)}_{\Delta}(z- i \epsilon)\right] \quad \text{for} \ z\in (1,\infty) \ .
\end{align}

\noindent We will now rewrite $I_{\Delta} $ above as an integral over the region $z\in (0,1)$ as done in \cite{Mazac:2018qmi}. For this, let us split the integral into three regions as follows:
\begin{align}
\label{euclOPEinva}
I_{\Delta} & = \int_{-\infty}^{0} dz z^{-2} \ \psi^{(-)}_{\Delta}(z) \mathcal{G}^{(-)}(z)+ \int_{0}^{1} dz z^{-2} \ \psi^{(0)}_{\Delta}(z) \mathcal{G}^{(0)}(z) \nonumber \\
& + \int_{1}^{\infty} dz z^{-2} \ \psi^{(+)}_{\Delta}(z) \mathcal{G}^{(+)}(z) \ . 
\end{align}

\noindent In the above expression, $\mathcal{G}^{(0)}(z)$ is the definition of the correlation function $\mathcal{G}(z)$ for $z\in (0,1)$, and  $\mathcal{G}^{(\pm)}(z)$ are given in terms of $\mathcal{G}^{(0)}(z)$ via \cite{Mazac:2018qmi},
\begin{align}
\label{relscorr}
& \mathcal{G}^{(-)}(z) = \mathcal{G}^{(0)}(z) \left(\frac{z}{z-1}\right) \quad \text{for} \ z\in (-\infty,0) \ ,\nonumber \\
& \mathcal{G}^{(+)}(z) = \pm z^{2\Delta_{\phi}}  \mathcal{G}^{(0)}\left(\frac{1}{z}\right) \quad \text{for} \ z\in (1,\infty) \ ,
\end{align}

\noindent where the $\pm$ sign in the last relation in Eq.~\eqref{relscorr} refers to the bosonic (fermionic) case. Then using the relations Eqs.~\eqref{relscpw},\eqref{relscorr} and employing a change of variables to recast all the integrals to the region $z \in (0,1)$, we can write the Euclident OPE inversion formula \eqref{euclOPEinv} as,
\begin{align}
\label{euclOPEinv1}
I_{\Delta} = \int_{0}^{1} dz z^{-2}\left[ 2 \psi^{(0)}_{\Delta}(z) \pm \frac{\psi^{(0)}_{\Delta}\left(\frac{1}{z}+ i\epsilon\right) +\psi^{(0)}_{\Delta}\left(\frac{1}{z}- i\epsilon\right) }{2 z^{2\Delta_{\phi}-2}} \right] \mathcal{G}^{(0)}(z). 
\end{align}

\noindent where again the relative $\pm$ sign in Eq.~\eqref{euclOPEinv1} refers to the bosonic (fermionic) case. This representation of $I_{\Delta}$ was previously obtained in \cite{Mazac:2018qmi}, which was useful for deriving the kernel in the Lorentzian OPE inversion formula.  

\vskip 4pt
 We can now insert the dispersion relation Eq.~\eqref{Dispersion_Relation} for $\mathcal{G}^{(0)}(z)$ in Eq.~\eqref{euclOPEinv1} to get,
\begin{align}
\label{euclOPEinv2}
I_{\Delta} & = \int_{0}^{1} dz z^{-2}\left[ 2 \psi^{(0)}_{\Delta}(z) \pm \frac{\psi^{(0)}_{\Delta}\left(\frac{1}{z}+ i\epsilon\right) +\psi^{(0)}_{\Delta}\left(\frac{1}{z}- i\epsilon\right) }{2 z^{2\Delta_{\phi}-2}} \right] \int_{0}^{1} dw \ K(\Delta_{\phi},z,w) \mathrm{dDisc}[\mathcal{G}^{(0)}(w)]  \nonumber \\
&=2  \int_{0}^{1} dw  w^{-2} \left(\int_{0}^{1} dz z^{-2}\left[ 2 \psi^{(0)}_{\Delta}(z) \pm \frac{\psi^{(0)}_{\Delta}\left(\frac{1}{z}+ i\epsilon\right) +\psi^{(0)}_{\Delta}\left(\frac{1}{z}- i\epsilon\right) }{2 z^{2\Delta_{\phi}-2}} \right]\frac{w^{2}}{2}K(\Delta_{\phi},z,w)\right) \times \nonumber \\
& \hspace{4.0cm} \times \mathrm{dDisc}[\mathcal{G}^{(0)}(w)].
\end{align}

\noindent where in the last equality we changed the order of integrations. Now Eq.~\eqref{euclOPEinv2} has the same form as the Lorentzian OPE inversion formula
\begin{align}
\label{lorOPEinv2}
I_{\Delta} & = 2  \int_{0}^{1} dw  w^{-2} \ H_{\Delta}(\Delta_{\phi},w) \mathrm{dDisc}[\mathcal{G}^{(0)}(w)] \ .
\end{align}

\noindent Comparing the previous two equations gives our desired formula:
\begin{align}
\label{lorOPEinvH}
H_{\Delta}(\Delta_{\phi},w)& = \int_{0}^{1} dz z^{-2}\left[ 2 \psi^{(0)}_{\Delta}(z) \pm \frac{\psi^{(0)}_{\Delta}\left(\frac{1}{z}+ i\epsilon\right) +\psi^{(0)}_{\Delta}\left(\frac{1}{z}- i\epsilon\right) }{2 z^{2\Delta_{\phi}-2}} \right]\frac{w^{2}}{2}K(\Delta_{\phi},z,w).
\end{align}

\noindent This relation shows that the inversion kernel $H_{\Delta}(\Delta_{\phi},w)$ is an integral transform of the conformal partial wave $\psi^{(0)}_{\Delta}(z) $ against the the dispersion kernel $K(\Delta_{\phi},z,w)$. A similar relation for dimensions $d\geq 2$ was derived in \cite{Carmi:2019cub}. For $\Delta_{\phi}=1$ and $\Delta_{\phi} =1/2$, using the expressions for the dispersion kernels given in Eq.~\eqref{KbDphi1} and Eq.~\eqref{KfDphihalf}, we have checked that the RHS of the relation \eqref{lorOPEinvH} indeed matches with the corresponding known expressions of the OPE inversion kernel given in \cite{Mazac:2018qmi}.


\section{Witten diagrams in $AdS_2$}
\label{sec:5}

Consider $AdS_2$ space-time, and a QFT in the bulk of $AdS_2$. For simplicity, we take the QFT to be bulk scalar fields with cubic and quartic couplings. The amplitudes on $AdS_2$ correspond to conformal correlators on the $1d$ boundary of $AdS$. For weak coupling, the AdS amplitudes can be evaluated perturbatively via Witten diagrams. In this section, we will derive explicit expressions for 4-point amplitudes, at tree-level and 1-loop order. Our main tool will be the spectral representation in AdS \cite{Costa:2014kfa} \cite{Giombi:2017hpr,Sleight:2020obc,Carmi:2018qzm,Carmi:2019ocp,Carmi:2021dsn,Carmi:2024tzj,Ankur:2023lum}.

\begin{figure}[t]
\centering
\includegraphics[clip,height=3.5cm]{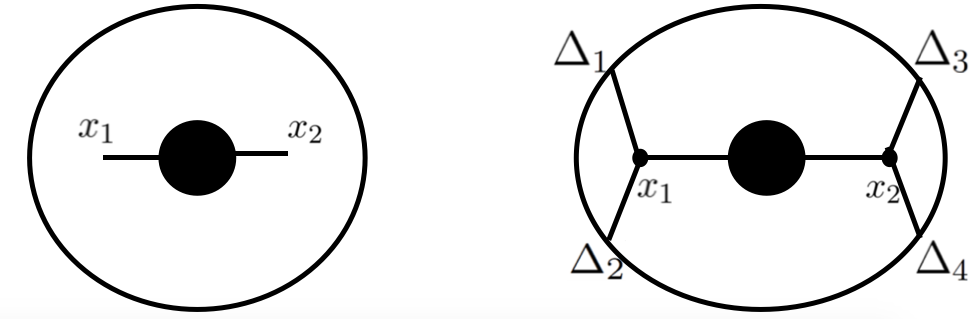}
\caption{Figure showing "Blob"diagrams for a scalar field in the bulk of AdS. \textbf{Left:} The black blob denotes a general bulk scalar two-point function $F(x_1,x_2)$. The blob is a general function of the geodesic distance in AdS. \textbf{Right:} Attaching 4 external legs to the boundary gives a conformal 4-point function. The "Blob diagrams'' are thus a class of diagrams which have two bulk-to-boundary propagators attached to $x_1$, and two attached to $x_2$.}
\label{fig:bubblerelations16}
\end{figure}

\subsection{The spectral representation}

Consider a general bulk 2-point function $F(x_1,x_2)$, of two points ($x_1$ and $x_2$) in the bulk, see Fig.~\ref{fig:bubblerelations16}-left. The function $F(x_1,x_2)$ can be expanded in a basis of AdS harmonic functions $\Omega_{\nu}(x_1,x_2)$:
\begin{align} 
\label{eq:nseeeeb}
F(x_1,x_2)\equiv \langle \phi(x_1) \phi(x_2) \rangle= \int_{-\infty}^\infty d\n  \tilde{F}_\n  \Omega _{\nu}(x_1,x_2) \,,
\end{align}
$\tilde{F}_\n$ is the spectral transform of $F(x_1,x_2)$, and $\n$ is the spectral parameter.

With the bulk 2-point function at hand, we can obtain a CFT correlator by attaching bulk-to-boundary propagators to the 2-point blob. For each bulk vertex one integrates over AdS space. As an example, consider the 4-point correlator (see Fig.~\ref{fig:bubblerelations16}-right),  obtained by attaching four bulk-to-boundary propagators:
\begin{align} 
\label{eq:jhg43}
&\langle \mm{O}_1(P_1)\cdots   \mm{O}_4(P_4)  \rangle = \int d^{d+1}x_2 d^{d+1}x_1  \times 
 \nn 
 &K_{\D_4}(P_4,x_2) K_{\D_3}(P_3,x_2)   K_{\D_2}(P_2,x_1) K_{\D_1}(P_1,x_1) F(x_1,x_2)   
\end{align} 
The $x_i$ are bulk points, and the $P_i$'s are boundary points. We call this family of AdS diagrams, shown in Fig.~\ref{fig:bubblerelations16}-right, ``Blob diagrams". Now we can plug Eq.~\eqref{eq:nseeeeb} inside Eq.~\eqref{eq:jhg43} and get:
\begin{align} 
\label{eq:sal23}
\langle \mm{O}_1(P_1) \ldots \mm{O}_4(P_4)  \rangle =  A \pi^2    g(z)
\end{align} 
where we defined:
\begin{align} 
\label{eq:nvjdf8}
A\equiv \l^2 \frac{  \Big( \frac{P_{24} }{P_{14}} \Big)^{a}  \Big( \frac{P_{14} }{P_{13}} \Big)^{b}  }{(P_{12})^{a_p }(P_{34})^{b_p}}
\end{align} 
and $a\equiv \frac{\D_1-\D_2}{2}$, $b\equiv \frac{\D_3-\D_4}{2}$ and $a_p\equiv \frac{\D_1+\D_2}{2}$, $b_p\equiv \frac{\D_3+\D_4}{2}$. As shown in e.g \cite{Costa:2014kfa,Carmi:2019ocp,Carmi:2021dsn,Carmi:2024tzj}, the function $g(z)$ in Eq.~\eqref{eq:sal23} is given by 
\begin{align} 
\label{eq:98d3}
 g(z)  = \int_{-\infty}^\infty d\nu 
\frac{ \tilde{F}_{\n}  }{  \sin \left[\pi (a_p-\frac{\frac{1}{2}+i \nu}{2})\right] \sin \left[\pi (b_p-\frac{\frac{1}{2}+i \nu}{2})\right]}  
 \Upsilon_{\nu}^{\D_i} \times \mathcal{K}^{\D_i}_{\frac{1}{2}+i\nu} (z) 
\end{align}
This equation is the expansion of the 4-point correlator in terms of the scalar conformal blocks $\mathcal{K}^{\D_i}_{\frac{d}{2}+i\nu} (z)$. The denominator $\sin\left[\pi (a_p-\frac{\frac{d}{2}+i \nu-l}{2})\right] \sin\left[\pi (b_p-\frac{\frac{d}{2}+i \nu-l}{2})\right]$ gives rise to poles that contribute to the double-trace operators. These double-trace poles are located at $\frac{d}{2}+i \nu = 2a_p+2n$ and $\frac{d}{2}+i \nu = 2b_p+2n$, where $n$ is an integer. Additional poles will come from the function $ \tilde{F}_{\n}$. The function $\Upsilon_{\nu}^{\D_i}$ in Eq.~\eqref{eq:98d3} is defined as\footnote{Throughout this section the notation $\Gamma_{x}$ stands for the gamma function $\Gamma(x)$.}:
\begin{align} 
\label{eq:sdskkf}
\Upsilon_{\nu}^{\D_i} \equiv \frac{\Bigg[ \frac{  \G_{a_p  -\frac{1}{4}+\frac{i \nu}{2}}  \G_{b_p  -\frac{1}{4}+\frac{i \nu}{2}}  }{\G_{1-a_p  +\frac{1}{4}+\frac{i \nu}{2}}  \G_{1-b_p  +\frac{1}{4}+\frac{i \nu}{2}}  }  \Bigg] \Big( \G_{a+\frac{i \nu+\frac{1}{2}}{2}} \G_{-a+\frac{i \nu+\frac{1}{2}}{2}} \G_{b+\frac{i \nu+\frac{1}{2}}{2}} \G_{-b+\frac{i \nu+\frac{1}{2}}{2}} \Big) }{\Gamma_{i\nu} \Gamma_{\frac{1}{2}+i\n}    } \ \ \ . 
\end{align} 
For more details on this formalism, see \cite{Carmi:2019ocp,Carmi:2021dsn,Carmi:2024tzj}.

\subsection{Blob diagrams with $a=0$ and $b=\frac{1}{4}$}

\begin{figure}[t]
\centering
\includegraphics[clip,height=3.5cm]{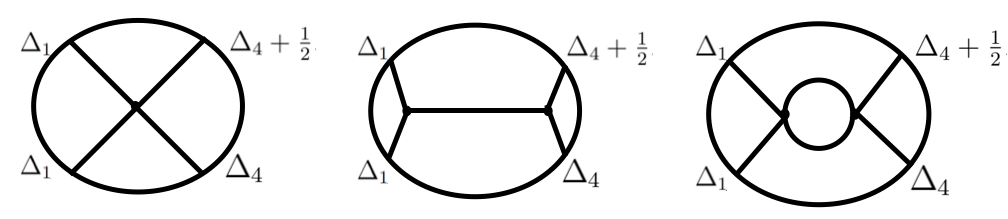}
\caption{Figure showing 4-point diagrams for a scalar field in the bulk of AdS with $a=0$ and $b=\frac{1}{4}$. \textbf{Left:} Contact diagram. \textbf{Middle:} Tree-level exchange diagram. \textbf{Right:} 1-loop bubble diagram.}
\label{fig:bubblerelations168}
\end{figure}

We will now apply the above formalism to a class of 4-point blob diagrams with external scaling dimensions\footnote{We are fixing just the differences of scaling dimension $\D_{12}$ and $\D_{34}$.} tuned to be $a\equiv \frac{\D_1-\D_2}{2} =0$ and $b\equiv \frac{\D_3-\D_4}{2} =\frac{1}{4}$. Examples are shown in Fig.~\ref{fig:bubblerelations168}. We will obtain explicit analytic expressions for these 4-point correlators. The higher dimensional cases $d=2,4$ were done in \cite{Carmi:2019ocp,Carmi:2021dsn}.

For scaling dimensions $a=0$ and $b=\frac{1}{4}$, the $d=1$ conformal blocks simplify from a ${}_2 F_1$ to a power law:
\bea
\label{eq:nds8d}
\mm{K}^{\D_i}_{\D}(z) = z^\D {}_2 F_1 (\D,\D+\frac{1}{2},2\D,z) = \sqrt{\frac{u}{v}} (2Z)^{\D-\frac{1}{2}}  
\eea
where we defined $Z\equiv \frac{2z}{(1+\sqrt{1-z})^2}$, and $u=z$ and $v=1-z$.  Using Eqs.~\eqref{eq:98d3} and \eqref{eq:sdskkf}, the spectral representation of a 4-point function \textit{Blob diagram} for the case of $a=0$ and $b=\frac{1}{4}$ is:
\begin{align} 
 g(z) = 4 \sqrt{2}  \pi  \int_{-\infty}^\infty d\nu \tilde{F}_{i\n} \frac{\Gamma_{\frac{1}{4}+\frac{i\n}{2} }  }{4^{i\n}\Gamma_{\frac{3}{4}+\frac{i\n}{2} }  } ( \G_{\D_1-\frac{1}{4}+\frac{i \nu}{2}}  \G_{\D_1-\frac{1}{4}-\frac{i \nu}{2}} \G_{\D_4+\frac{1}{4}+\frac{i \nu}{2}} \G_{\D_4+\frac{1}{4}-\frac{i \nu}{2}})  \mathcal{K}^{\D_i}_{\frac{1}{2}+i\n}(z)
\end{align}
Here $\tilde{F}_{i\n}$ is the spectral representation of the 2-point  ``blob" that appears in the diagram. For the moment it is a completely general function of $\n$. Using Eq.~\eqref{eq:nds8d}, and defining $\hat{g}(z) \equiv \frac{g(z)}{4 \sqrt{2} \pi  \sqrt{\frac{u}{v}}}$ gives:
\begin{align} 
\label{eq:raquel}
\hat{g}(z)=   \int_{-\infty}^\infty d\nu \tilde{F}_{i\n} \frac{\Gamma_{\frac{1}{4}+\frac{i\n}{2} }  }{ \Gamma_{\frac{3}{4}+\frac{i\n}{2} }  } ( \G_{\D_1-\frac{1}{4}+\frac{i \nu}{2}}  \G_{\D_1-\frac{1}{4}-\frac{i \nu}{2}} \G_{\D_4+\frac{1}{4}+\frac{i \nu}{2}} \G_{\D_4+\frac{1}{4}-\frac{i \nu}{2}})  Z^{i\n}
\end{align}
Using the identity $\G_x\G_{1-x} =\frac{\pi}{\sin (\pi x)}$, we can write this in a way which shows the double-trace poles explicitly:
\begin{align} 
\label{eq:poldhhd}
\hat g(z) =  \pi^2  \int_{-\infty}^\infty  d\nu  \frac{  \tilde{F}_{i\n}   }{\sin \left[\pi (\frac{2\D_1-\frac{1}{2}-i\n}{2})\right] \sin \left[\pi (\frac{2\D_4+\frac{1}{2}-i\n}{2})\right]}  \bigg( \frac{\Gamma_{\frac{1}{4}+\frac{i\n}{2} }  }{ \Gamma_{\frac{3}{4}+\frac{i\n}{2} }  } \frac{\G_{\D_1-\frac{1}{4}+\frac{i \nu}{2}}  }{\G_{-\D_1+\frac{5}{4}+\frac{i \nu}{2}}}  \frac{ \G_{\D_4 +\frac{1}{4}+\frac{i \nu}{2}}  }{\G_{\frac{3}{4}-\D_4 +\frac{i \nu}{2}}} \bigg) Z^{i\n}  
\end{align}
Closing the contour and using the residue theorem, we have 2 types of poles. There are possible poles from $\tilde{F}_{i\n}$, whose contribution we denote by $\hat g_{F}(z)$. There are also double-trace poles coming from the two $\sin$ factors in the denominator, whose contribution we denote by $\hat g_{d.t}(z)$. All together,
\begin{align} 
\label{eq:dfkhdf}
\hat g(z) = \hat g_{d.t}(z) +\hat g_{F}(z)
\end{align} 
where the double-trace contribution is:
\begin{align} 
\label{eq:jdfbbdjfdf}
&\hat g_{d.t}(z) = \frac{2\pi^2}{\cos (\pi \D_{14})} \sum_{m=0}^\infty   \frac{\G_{\D_1+\D_4  +m }}{\G_{\D_4-\D_1+\frac{3}{2} +m}} \frac{\G_{2\D_4+m+\frac{1}{2}} \G_{\D_4+m+\frac{1}{2}} }{\G_{\D_4+m+1} \G_{m+1}} \tilde{F}_{ 2\D_4+2m+\frac{1}{2}}   Z^{ 2\D_4+\frac{1}{2}+2m} 
\nn
&+ \Big( \D_1 \leftrightarrow \D_4+\frac{1}{2} \Big)
\end{align}
One can compute the double discontinuity of the 4-point function by acting with $\mathrm{dDisc}$ on both sides of Eq.~\eqref{eq:poldhhd}. The double discontinuity of the conformal block cancels the $\sin$ factors in Eq.~\eqref{eq:poldhhd}, and gives:
\begin{align} 
\label{eq:poldhhd2}
\mathrm{dDisc} [\hat g(z) ]=  \mathrm{dDisc} [\hat g_{F}(z)] =  \pi^2  \int_{-\infty}^\infty  d\nu   \tilde{F}_{i\n}   \bigg( \frac{\Gamma_{\frac{1}{4}+\frac{i\n}{2} }  }{ \Gamma_{\frac{3}{4}+\frac{i\n}{2} }  } \frac{\G_{\D_1-\frac{1}{4}+\frac{i \nu}{2}}  }{\G_{-\D_1+\frac{5}{4}+\frac{i \nu}{2}}}  \frac{ \G_{\D_4 +\frac{1}{4}+\frac{i \nu}{2}}  }{\G_{\frac{3}{4}-\D_4 +\frac{i \nu}{2}}} \bigg) Z^{i\n}  
\end{align}
In particular, we see that taking the double discontinuity cancels the double-trace contribution.

In the following subsections, we will use the above formulae to explicitly compute tree-level and 1-loop diagrams in positions space, i.e. as a function of the cross-ratio $z$.

\subsubsection{Tree-level Contact diagram}
\label{sec:contdi}
The contact diagram, Fig.~\ref{fig:bubblerelations168}-left, has a spectral representation which is a constant, i.e. $\tilde{F}_\n =1$. 
Therefore $\hat g_{F}(z)$ in Eq.~\eqref{eq:dfkhdf} is zero. The double-trace contribution from Eq.~\eqref{eq:jdfbbdjfdf} is:
\begin{align} 
&\hat g(z)= \hat g_{d.t}(z) = \frac{2\pi^2}{\cos (\pi \D_{14})} \sum_{m=0}^\infty   \frac{\G_{\D_1+\D_4  +m }}{\G_{\D_4-\D_1+\frac{3}{2} +m}} \frac{\G_{2\D_4+m+\frac{1}{2}} \G_{\D_4+m+\frac{1}{2}} }{\G_{\D_4+m+1} \G_{m+1}}  Z^{ 2\D_4+\frac{1}{2}+2m} 
\nn
&+ \Big( \D_1 \leftrightarrow \D_4+\frac{1}{2} \Big)
\end{align}
This sum is simply the Taylor series of a hypergeometric function:
\begin{align}
\label{eq:789s1}
&\hat g(z) =  \frac{2\pi^2}{\cos (\pi \D_{14})}  \G_{2\D_4+\frac{1}{2}} \G_{\D_4+\frac{1}{2}}  \G_{\D_1+\D_4} Z^{2 \D_4+\frac{1}{2}} \times
\nn
&\ {}_3 F^{(reg)}_2 (\D_4+\frac{1}{2}, \D_1+\D_4 ,2\D_4+\frac{1}{2}:\D_4+1, \D_4-\D_1+\frac{3}{2}, Z^{2})
+ \Big( \D_1 \leftrightarrow \D_4+\frac{1}{2} \Big)
\end{align}
where ${}_3 F^{(reg)}_2(a,b,c:d,e,x) \equiv \frac{1}{\G_d \G_e}{}_3 F_2(a,b,c:d,e,x)$ is the regularized hypergeometric function. Eq.~\eqref{eq:789s1} is the result for the 1d contact diagrams with scaling dimensions $\D_2=\D_1$ and $\D_3=\D_4+\frac{1}{2}$.








\subsubsection{Tree-level Exchange diagram}
Consider the exchange diagram, Fig.~\ref{fig:bubblerelations168}-middle, with the exchange of an operator of dimension $\D$. The spectral representation in this case is $\tilde{F}_{i\n} = \frac{1}{\n^2+(\D-\frac{1}{2})^2}$. Plugging this in Eq.~\eqref{eq:raquel} gives:
\begin{align} 
\hat g(z)=  \int_{-\infty}^\infty d\nu  \frac{ \frac{\Gamma_{\frac{1}{4}+\frac{i\n}{2} }  }{ \Gamma_{\frac{3}{4}+\frac{i\n}{2} }  } ( \G_{\D_1-\frac{1}{4}+\frac{i \nu}{2}}  \G_{\D_1-\frac{1}{4}-\frac{i \nu}{2}} \G_{\D_4+\frac{1}{4}+\frac{i \nu}{2}} \G_{\D_4+\frac{1}{4}-\frac{i \nu}{2}}) }{\n^2+(\D-\frac{1}{2})^2}  Z^{i\n}  
\end{align}
There is one pole from the exchange operator at $i\n=\D-\frac{1}{2}$, which gives:
\begin{align} 
\hat g_{F}(z)=   \frac{\pi}{(\D-\frac{1}{2})} \frac{ \G_{\frac{\D}{2} } \G_{\D_1-\frac{\D}{2}} \G_{\D_1+\frac{\D}{2}-\frac{1}{2}} \G_{\D_4 +\frac{1}{2} -\frac{\D}{2}} \G_{\D_4 +\frac{\D}{2}}}{\G_{\frac{\D}{2} +\frac{1}{2}} }  Z^{\D-\frac{1}{2}}
\end{align}
The contribution from the double-traces is:
\begin{align} 
\label{eq:slkjd3}
&\hat g_{d.t}(z ) =\frac{2 \pi^2 }{\cos (\pi \D_{14})} \sum_{m=0}^\infty   \frac{\frac{\G_{\D_1+\D_4  +m }}{\G_{\D_4-\D_1+\frac{3}{2} +m}} \frac{\G_{2\D_4+m+\frac{1}{2}} \G_{\D_4+m+\frac{1}{2}} }{\G_{\D_4+m+1} \G_{m+1}}  }{-(2\D_4+2m+\frac{1}{2})^2+(\D-\frac{1}{2})^2}  Z^{ 2\D_4+\frac{1}{2}+2m} 
\nn
&+ \Big( \D_1 \leftrightarrow \D_4+\frac{1}{2} \Big)
\end{align}
The result of the sum is a hypergeometric ${}_4F_3$:
\begin{align} 
&\hat g_{d.t}(z) = 2^{-6\D_4}\pi^3 \frac{\G_{4\D_4+1}   \G_{\D_1+\D_4}Z^{2\D_4+\frac{1}{2}}}{(2\D-1)\Gamma_{\D_4+1}\cos \pi(\D_1-\D_4)}
 \nn
&\times \bigg[ {\scriptstyle  \G_{\D_4+\frac{\D}{2} }  {}_4 F_3^{(reg)} (\D_4+\frac{1}{2},\D_4+\frac{\D}{2} ,\D_1+\D_4,2\D_4+\frac{1}{2}:\D_4+1,\D_4+\frac{\D}{2}+1,\frac{3}{2}+\D_4-\D_1,Z^{2})}
\nn
& - {\scriptstyle \G_{\D_4-\frac{\D}{2}+\frac{1}{2}}{}_4 F_3^{(reg)} (\D_4+\frac{1}{2},\D_4-\frac{\D}{2}+\frac{1}{2},\D_1+\D_4,2\D_4+\frac{1}{2}:\D_4+1,\D_4-\frac{\D}{2}+\frac{3}{2},\frac{3}{2}+\D_4-\D_1,Z^{2}) } \bigg]
\nn
&+ \Big( \D_1 \leftrightarrow \D_4+\frac{1}{2} \Big)
\end{align}
Thus, the result for this family of exchange diagrams is $\hat g(z) = \hat g_{d.t}(z) +\hat g_{F}(z)$. 

\vskip 4pt
Using the above results for the contact and exchange Witten diagrams, we can compute the $d=1$ bosonic Polyakov blocks for external scaling dimensions satisfying $\D_1 =\D_{2}$ and $\D_3=\D_4+1/2$. Polyakov blocks are defined as crossing symmetric combinations of tree-level exchange and contact Witten diagrams. Here we have evaluated the exchange Witten diagram only in a particular channel. To obtain the Polyakov block, we will also need to take into account the other two exchange channels. Polyakov blocks for $1$-$d$ CFT correlators, where external operators have unequal scaling dimensions, have been recently obtained in \cite{Ghosh:2023wjn}. Explicit expressions for Polyakov blocks for identical external operators with $\Delta_{\phi}=1$ and $\Delta_{\phi}=1/2$ have been derived in \cite{Bonomi:2024lky} using the representation of Polyakov blocks as an integral of the double-discontinuity of the conformal block times the $1$-$d$ CFT dispersion relation kernel \cite{Paulos:2020zxx,Bonomi:2024lky}. 

\subsubsection{1-loop bubble diagram}
We will now consider the 1-loop bubble diagram, Fig.~\ref{fig:bubblerelations168}-right, with the bulk propagator in the bubble having a scaling dimension $\D$. We will denote the spectral representation of the 1-loop bubble bulk 2-point functions as $F_{i\n} = \tilde{B}_\n$.
The double-discontinuity of this diagram, from Eq.~\eqref{eq:poldhhd2} is:
\begin{align} 
\label{eq:pksdbfs9}
\mathrm{dDisc}[\hat g(z)] =    \int  d\nu    \tilde{B}_{\n}     \frac{\Gamma_{\frac{1}{4}+\frac{i\n}{2} }  }{ \Gamma_{\frac{3}{4}+\frac{i\n}{2} }  } \frac{\G_{\D_1-\frac{1}{4}+\frac{i \nu}{2}}  }{\G_{-\D_1+\frac{5}{4}+\frac{i \nu}{2}}}  \frac{ \G_{\D_4 +\frac{1}{4}+\frac{i \nu}{2}}  }{\G_{\frac{3}{4}-\D_4 +\frac{i \nu}{2}}} Z^{i\n}  
\end{align}
The poles of $\tilde{B}_\n$ were computed in \cite{Carmi:2018qzm} for $AdS_{d+1}$, with the result:
\begin{align}
2\tilde{B}_\n \sim \frac{-1}{\frac{d}{2} + i\n-(2\D+2m)} \frac{(\frac{d}{2})_m \G_{\D+m} \G_{\D+m-\frac{d}{2}+\frac{1}{2}} \G_{2\D+m-\frac{d}{2}} }{(4\pi)^d \G_{m+1} \G_{\D+m+\frac{1}{2}} \G_{\D+m-\frac{d}{2}+1} \G_{2\D+m-d+1}}
\end{align}
Plugging this result in Eq.~\eqref{eq:pksdbfs9} and  taking $d=1$ we get:
\begin{align} 
&\mathrm{dDisc}[\hat g(z)] = -  \frac{1}{4}  \sum_{m=0}^\infty \frac{(\frac{1}{2})_m \G_{m+\D}^3 \G_{m+2\D-\frac{1}{2}}  \G_{-\frac{1}{2}+m+\D+\D_1} \G_{m+\D +\D_4}}{ \G_{m+1} \G^3_{\D+m+\frac{1}{2}} \G_{m+2\D} \G_{1+m+\D-\D_1}   \G_{\frac{1}{2}+m+\D-\D_4} }   Z^{2m+2\D-\frac{1}{2}}
\nn
&= -\frac{1}{4}  \G_\D^3 \G_{2\D-\frac{1}{2}} \G_{\D+\D_1-\frac{1}{2}} \G_{\D+\D_4 }Z^{2\D-\frac{1}{2}}  \times
\nn
& {}_7 F_6 ({ \scriptstyle \frac{1}{2}, \D,\D,\D,2\D-\frac{1}{2},\D_1+\D-\frac{1}{2},\D_4+\D:
2\D, \frac{2\D+1}{2}+\D, \frac{1}{2}+\D, \frac{1}{2}+\D,1+\D-\D_1,\frac{1}{2}+\D-\D_4 ,Z^{2})}
\end{align}



\section{Conclusions}
\label{sec:concl}

In this work, we have obtained a dispersion relation for $d=1$ CFT four-point correlation function of identical operators with scaling dimension $\Delta_{\phi}$ being integers (for bosons) and half-integers (for fermions). We have also studied tree-level and $1$-loop $AdS_{2}$ Witten diagrams for bulk scalars with cubic and quartic interactions. 

An important direction for future work would be to derive analytical formulas for the $1$-$d$ dispersion kernel for general values of $\Delta_{\phi}$. One method towards this would be to find the kernel in the Lorentzian inversion formula for general $\Delta_{\phi}$ and then use that to compute the dispersion kernel following the same procedure discussed in this paper. Another method of computing the dispersion kernel is to solve the integral equation \eqref{gwidehatPaulos} for the kernel found in \cite{Paulos:2020zxx}. Both approaches should yield the same result. The integral equation in \cite{Paulos:2020zxx} can be solved numerically for any $\Delta_{\phi}$, but it would be desirable to find analytical methods to solve it for general $\Delta_{\phi}$. 

It would be interesting to derive the Lorentzian inversion formula kernel and the dispersion kernel for mixed correlators\footnote{Dispersion relation for mixed correlators in $d\ge 2$ CFTs will appear in \cite{carmi}.} which involve external operators with unequal scaling dimensions. A $1$-$d$ CFT dispersion relation for mixed correlators would be useful for providing an alternative formulation of the $1$-$d$ CFT mixed correlator Polyakov bootstrap which has recently been initiated in \cite{Ghosh:2023wjn}. 

In \cite{Carmi:2019cub}, a direct proof of the dispersion relation for $d\ge 2$ CFTs was given using a contour deformation argument. It would be worth knowing if such a contour method can be used to derive the dispersion relation in the current paper.

In \cite{Paulos:2020zxx}, it is shown that all constraints from the crossing equation in one-dimensional CFTs (or along the $z=\bar{z}$ line in higher-dimensional CFTs) can be equivalently formulated through a variety of approaches, including the crossing-symmetric dispersion relation, the Polyakov bootstrap, or sum rules derived from a complete basis of analytic functionals that act on the crossing equation. In particular, given the dispersion relation on the line one can compute a complete basis of extremal analytic functionals, called master functionals. These functionals can, in principle, be used to construct sum rules and thus impose constraints on the CFT data, highlighting the importance of computing the dispersion relation. An interesting problem would be to compute lower bounds on the central charge \cite{Rattazzi:2010gj,Gadde:2023daq}, or equivalently, upper bounds on the stress-tensor OPE coefficient squared, $c^2_{\phi \phi T_{\mu \nu}}$\footnote{See \cite{Mazac:2018mdx} for related work in the large $\Delta_{\phi}$ limit.}, by constructing suitable functionals and making assumptions about the low-lying spectrum (such as a gap). Through AdS/CFT, this will give the upper bounds on the strength of gravity in the dual bulk theory in AdS.

Studying large $\Delta_{\phi}$ limit of correlators in $d$ dimensional CFTs is significant as it is related to the scattering amplitudes of massive particles in quantum field theories in $d+1$ dimensional Minkowski spacetime \cite{Paulos:2016fap,Dubovsky:2017cnj,Hijano:2019qmi,Komatsu:2020sag,Gadde:2022ghy,Cordova:2022pbl,vanRees:2022zmr,vanRees:2023fcf}. In \cite{Paulos:2020zxx}, it was shown that, in this limit, the integral equation \eqref{gwidehatPaulos} for the dispersion kernel in one dimension reduces to an algebraic equation, making the computation of the dispersion kernel much easier. A natural extension of this would be to derive perturbative corrections in $1/\Delta_{\phi}$ to this kernel, using one of the two previously discussed methods, to probe the effects of AdS curvature corrections. A potential challenge in this approach is that the large $\Delta_{\phi}$ limit does not commute with the small cross-ratio limit, which may introduce some complications in the analysis. 

The $1$-$d$ conformal dispersion relation in the current paper and the higher-$d$ dispersion relation \cite{Carmi:2019cub}, are both written in position space. Conformal correlators are often naturally expressed in Mellin space \cite{Mack:2009mi,Penedones:2010ue,Fitzpatrick:2011ia}, and the Mellin amplitude has properties reminiscent of scattering amplitudes in Minkowski space. The Mellin amplitude obeys a Cauchy-type dispersion relation \cite{Penedones:2019tng,Carmi:2020ekr}, of the form Eq.~\eqref{eq:ds4}. It was then shown in \cite{Caron-Huot:2020adz} that the Mellin space dispersion relation is the Mellin transform of the position space dispersion relation. It would be interesting to understand if one can find a similar relation for the case of the $1$-$d$ dispersion relation in the current paper, for example by using the definition of Mellin amplitudes for $1$-$d$ CFTs given in \cite{Bianchi:2021piu}. 

\section*{Acknowledgments}

We thank Miguel Paulos, Davide Bonomi and Valentina Forini for helpful correspondence. We also thank the CERN theory group and CERN, where part of this project was done. This work was supported by the Israeli Science Foundation (ISF) grant number 1487/21 and by the MOST NSF/BSF physics grant number 2022726.


\appendix

\section{Dispersion kernel computations}\label{Dispersion_Kernel}

\subsection{Computing $K^{(p)}_{princ}$}\label{K_princ_p_Appen}

In this section we will compute the kernel $K^{(p)}_{princ}$ by evaluating the $\Delta$ integral in Eq.~\eqref{eq:nb3}. This integral can be evaluated by closing the contour and using Cauchy's residue theorem. When $z,w <1$, we should close the contour to the right because the integrand decays at $Re(\D) \to +\infty$, and we can drop the arc at infinity. Now let us look at the poles of the integrand of Eq.~\eqref{eq:nb3}. There are double poles from the $\sin^2[\frac{\pi \D}{2}]$ denominator, at $\D= 2m+2$, for $m=0,1,2, \ldots$. After picking up these poles, the result of the $\Delta$ integral is
\begin{align}
\label{eq:plv2}
K^{(p)}_{princ}(\Delta_{\phi},z,w) &=  - \frac{2}{\pi^{2}w^2} \Bigg(w^{2-2\D_{\phi}}  \sum_{m=0}^\infty   \frac{d}{dm'} \big[ (4m'+3)P_{2m'+1}(\tilde w) Q_{2m'+1}(\hat z) \big] \big|_{m'=m} \nonumber \\
& \hspace{1.5cm}+\Big(w \to \frac{w}{w-1} \Big)  \Bigg) \ .
\end{align}

\noindent where $\tilde{w}=1-2w$ and $\hat{z}= \frac{2}{z}-1$. The result of this sum, for any integer $\Delta_{\phi}$, is computed in Eq.~\eqref{sumdrPQ},
\begin{align}
K^{(p)}_{princ}(\Delta_{\phi},z,w) & = -\frac{2}{\pi^{2}w^{2}} \Bigg( w^{2-2\D_{\phi}}  \Big(
\frac{1}{\tilde w-\hat z} \log \Big[\frac{\hat z+1}{\tilde w+1}\Big]  + \frac{1}{\tilde w+\hat z} \log \Big[\frac{\hat z-1}{\tilde w+1}\Big]  \Big) \nonumber \\
& \hspace{1.5cm}+\Big(w \to \frac{w}{w-1} \Big) \Bigg) \ .
\end{align}

\noindent In terms of the $z$ and $w$ coordinates the above expression becomes,
\begin{align}\label{Kpzw}
 K^{(p)}_{princ}(\Delta_{\phi},z,w) = \frac{z}{\pi^2w }   \bigg[   w^{1-2\D_\phi}    \bigg(
\frac{\log \big(\frac{1-z}{z(1-w)}\big)}{w z-1}  - \frac{\log \big( z(1-w) \big)}{1+ z(w-1)}  \bigg) 
\nn
+\Big(\frac{w}{w-1}\Big)^{1-2\D_{\phi}}    \bigg(
\frac{\log \big(\frac{(1-w)(1-z)}{z}\big)}{1-w(1-z)}  - \frac{\log \big(  \frac{1-w}{z} \big)}{1-w-z}  \bigg)  \bigg].
\end{align}

\subsection{Computing $K^{(q)}_{princ}$}
\label{app:Kqprinc}

Here we compute the $q$-term in the principal series contribution to the dispersion kernel. This term is given by Eq.~\eqref{eq:nb4},
\begin{align}
\label{Kqcomp}
K^{(q)}_{princ}(\Delta_{\phi},z,w)= 
  w^{-2 }   \int_{\frac{1}{2}-i\infty}^{\frac{1}{2}+i\infty} \frac{d \D}{2\pi i}  \   \frac{(2\D-1)}{\sin^2[\frac{\pi \D}{2}]} \  q_\D^{\D_\phi}(w) Q_{\D-1}(\hat z) \ .
\end{align}

\noindent where $\hat{z}=\frac{2}{z}-1$. We will now consider the bosonic and fermionic cases separately as follows. 

\subsubsection*{Bosonic Case}

In this case, $q_\D^{\D_\phi}(w)$ for $\D_{\phi} \in \mathbb{N}$ is given by
\begin{align}
\label{qb}
q_\D^{\D_\phi}(w) = \frac{a^{\Delta_{\phi}}_{\Delta}(w)+b^{\Delta_{\phi}}_{\Delta}(w)\log(1-w)}{w^{2\Delta_{\phi}-2} }
\end{align}

\noindent where $a^{\Delta_{\phi}}_{\Delta}(w), b^{\Delta_{\phi}}_{\Delta}(w)$ are polynomials in $\Delta$ and $w$ and can be expressed as
\begin{align}
\label{qb1}
a^{\Delta_{\phi}}_{\Delta}(w) = \sum_{p=0}^{2\Delta_{\phi}-4} \widetilde{a}^{\Delta_{\phi}}_{p}(w) \widetilde{\Delta}^{p}, \quad b^{\Delta_{\phi}}_{\Delta}(w) = \sum_{p=0}^{2\Delta_{\phi}-4}  \widetilde{b}^{\Delta_{\phi}}_{p}(w) \widetilde{\Delta}^{p}
\end{align}

\noindent where we have defined $\widetilde{\D} = \D(\D-1)$. For example \cite{Mazac:2018qmi},
\begin{align}
\label{qbexamples}
& \Delta_{\phi}=1: \quad  a^{1}_{\Delta}(w) = 0, \quad b^{1}_{\Delta}(w) = 0 \nonumber \\
& \Delta_{\phi}=2: \quad  a^{2}_{\Delta}(w) = w^{2}+2w-2, \quad b^{2}_{\Delta}(w) = 0 \nonumber \\
& \Delta_{\phi}=3: \quad  a^{3}_{\Delta}(w) = w^{4}+ \frac{1}{2}(\widetilde{\D}^{2}-8\widetilde{\D}+8)w^{3}- \frac{1}{2}(\widetilde{\D}-2)(\widetilde{\D}-6)w^{2}+4w-2, \nonumber \\
& \hspace{1.7cm} \quad b^{3}_{\Delta}(w) = 0.
\end{align}

\noindent In contrast to the first few cases shown above, $b^{\Delta_{\phi}}_{\Delta}(w)$ is non-zero for $\Delta_{\phi} \ge 4$. Now using \eqref{qb} in \eqref{Kqcomp} we get
\begin{align}
\label{Kqa}
K^{(q)}_{princ}(\Delta_{\phi},z,w) & = 
  w^{-2\Delta_{\phi}} \sum_{p=0}^{2\Delta_{\phi}-4}  \left(\widetilde{a}^{\Delta_{\phi}}_{p}(w) + \widetilde{b}^{\Delta_{\phi}}_{p}(w) \log(1-w)\right) \mathcal{I}_{p}(z) 
\end{align}

\noindent where $\mathcal{I}_{p}(z) $ is the following integral
\begin{align}
\label{Ip}
 \mathcal{I}_{p}(z)  & =   \int_{\frac{1}{2}-i\infty}^{\frac{1}{2}+i\infty} \frac{d \D}{2\pi i}  \   \frac{(2\D-1)}{\sin^2[\frac{\pi \D}{2}]} \  \widetilde{\Delta}^{p} \ Q_{\D-1}(\hat z) \ .
\end{align}

\noindent To evaluate the above integral, we note the identity
\begin{align}
\label{PJdiffn}
& \mathcal{D}_{w}^{p} P_{\Delta-1}(1-2w)\big|_{w=0}= \widetilde{\Delta}^{p}, \quad \mathcal{D}_{w} \equiv \partial_{w}\left[(w^{2}-w)\partial_{w}\right].
\end{align} 

\noindent Then we have,
\begin{align}
\label{Ip1}
 \mathcal{I}_{p}(z)   & = \mathcal{D}_{w}^{p} \left[ \int_{\frac{1}{2}-i\infty}^{\frac{1}{2}+i\infty} \frac{d \D}{2\pi i}   \frac{(2\D-1)}{\sin^2[\frac{\pi \D}{2}]}  \ P_{\Delta-1}(1-2w) Q_{\D-1}(\hat z) \right] \Bigg|_{w=0}.
\end{align}

\noindent Thus we can evaluate $\mathcal{I}_{p}(z)$ by acting with the differential operator $\mathcal{D}_{w}^{p}$ on the following integral
\begin{align}
\label{Ip2}
 &  \int_{\frac{1}{2}-i\infty}^{\frac{1}{2}+i\infty} \frac{d \D}{2\pi i}    \frac{(2\D-1)}{\sin^{2}\left(\frac{\pi\Delta}{2}\right)} \ P_{\Delta-1}(1-2w) Q_{\D-1}(\hat z)  \nonumber \\
 & = \frac{z}{\pi^2  }     \bigg(
\frac{\log \big(\frac{1-z}{z(1-w)}\big)}{w z-1}  - \frac{\log \big( z(1-w) \big)}{1+ z(w-1)} \bigg) 
\end{align}

\noindent where we have applied the identity in Eq.~\eqref{sumdrPQ} to compute the integral. Using Eq.~\eqref{Ip2} in Eq.~\eqref{Ip1} we can then compute the integral $ \mathcal{I}_{p}(z)$. We note below the explicit expressions for $K^{(q)}_{princ}(\Delta_{\phi},z,w) $ for $\Delta_{\phi}=1,2,3$. 
\begin{align}
\label{Kqb1}
          &   K^{(q)}_{princ}(\Delta_{\phi}=1,z,w) =0 \\ 
          \label{Kqb2}
          & K^{(q)}_{princ}(\Delta_{\phi}=2,z,w)= - \frac{ z (w^{2}+2 w-2)}{\pi^{2}w^{4}} \left[\frac{z}{1-z} \log(z)+ \log(1-z)\right]\\
          \label{Kqb3}
           & K^{(q)}_{princ}(\Delta_{\phi}=3,z,w)= - \frac{ z (w^{4}+4w^{3}-6w^{2}+4w-2)}{\pi^{2}w^{6}} \left[\frac{z}{1-z} \log(z)+ \log(1-z)
\right]\nonumber \\
& \hspace{0.5cm}+ \frac{ z^{2} (1-w)}{\pi^{2}w^{4}} \left[\frac{2z^{2}}{(1-z)^2}+ \frac{(2z^4-3z^3-3 z^{2}+12 z-6)}{(1-z)^3}\log(z) +(2 z+3)\log(1-z)\right].
\end{align} 

\subsubsection*{Fermionic case}

We will now consider the fermionic case  and take $\D_{\phi} \in \mathbb{N} +\frac{1}{2}$. Let us denote the $q$-term in the Lorentzian inversion kernel as $\widehat{q}_\D^{\D_\phi}(w)$. This again takes a form similar to Eq.~\eqref{qb},
\begin{align}
\label{qbf}
\widehat{q}_\D^{\D_\phi}(w) = \frac{\widehat{a}^{\Delta_{\phi}}_{\Delta}(w)+\widehat{b}^{\Delta_{\phi}}_{\Delta}(w)\log(1-w)}{w^{2\Delta_{\phi}-2} }
\end{align}

\noindent where $\widehat{a}^{\Delta_{\phi}}_{\Delta}(w), \widehat{b}^{\Delta_{\phi}}_{\Delta}(w)$ are polynomials in $\Delta$ and $w$ and can be written as,
\begin{align}
\label{qf}
\widehat{a}^{\Delta_{\phi}}_{\Delta}(w) = \sum_{p=0}^{2\Delta_{\phi}-2} \widetilde{\mathtt{a}}^{\Delta_{\phi}}_{p}(w) \widetilde{\Delta}^{p}, \quad \widehat{b}^{\Delta_{\phi}}_{\Delta}(w) = \sum_{p=0}^{2\Delta_{\phi}-2}  \widetilde{\mathtt{b}}^{\Delta_{\phi}}_{p}(w) \widetilde{\Delta}^{p}
\end{align}

\noindent For example we have \cite{Mazac:2018qmi},
\begin{align}
\label{qfexamples}
& \Delta_{\phi}=1/2: \quad  \widehat{a}^{1/2}_{\Delta}(w) = 0 \ , \quad \widehat{b}^{1/2}_{\Delta}(w) = 0 \nonumber \\
& \Delta_{\phi}=3/2: \quad  \widehat{a}^{3/2}_{\Delta}(w) = (2\widetilde{\Delta}-1)w, \quad \widehat{b}^{3/2}_{\Delta}(w) = 0 \nonumber \\
& \Delta_{\phi}=5/2: \quad  \widehat{a}^{5/2}_{\Delta}(w) = \frac{1}{105}(2\widetilde{\D}^{3}-25\widetilde{\D}^{2}+216\widetilde{\D}-243)w(w^{2}-w+1) \ , \nonumber \\
& \hspace{2.5cm} \widehat{b}^{5/2}_{\Delta}(w) = \frac{1}{210}(\widetilde{\D}-2)(\widetilde{\D}-12)(2\widetilde{\D}+3)(w-2)(2w^{2}+w-1) \ .
\end{align}

\noindent The $q$-term of the principal series kernel then takes the general form
\begin{align}
\label{Kqa}
\widehat{K}^{(q)}_{princ}(\Delta_{\phi},z,w) & = 
  w^{-2\Delta_{\phi}} \sum_{p=0}^{2\Delta_{\phi}-2}  \left(\widetilde{\mathtt{a}}^{\Delta_{\phi}}_{p}(w) + \widetilde{\mathtt{b}}^{\Delta_{\phi}}_{p}(w) \log(1-w)\right) \mathcal{I}_{p}(z) 
\end{align}

\noindent where $\mathcal{I}_{p}(z)$ is the same integral as in the bosonic case given in Eq.~\eqref{Ip}. We provide below the explicit results for $\Delta_{\phi}=1/2,3/2,5/2$.
\begin{align}
\label{Kqf1}
          &   \widehat{K}^{(q)}_{princ}(\Delta_{\phi}=1/2,z,w) =0 \\
          \label{Kqf3by2}
          &  \widehat{K}^{(q)}_{princ}(\Delta_{\phi}=3/2,z,w)=\frac{ z }{\pi^{2}w^{2}} \left[\frac{2z}{z-1}+\frac{z(z(3-2z)-3)}{(1-z)^{2}} \log(z)+(2z+1)\log(1-z)\right] \\
          &  \widehat{K}^{(q)}_{princ}(\Delta_{\phi}=5/2,z,w)= \frac{ z (w^{2}-w+1)}{35\pi^{2}w^{4}} \bigg[\frac{24  z(z^{2}-z+1)^2}{ (z-1)^3} \nonumber \\
         & \hspace{2.5cm}- \frac{ z(24 z^6-84 z^5+154 z^4-175 z^3+175 z^2-105 z+35) }{ (z-1)^4} \log (z) \nonumber \\ 
        & \hspace{2.5cm} +  (24z^{3}+12z^{2}+58 z+81) \log (1-z)
\bigg]\nonumber \\
& \hspace{2.5cm}+\frac{ z (w-2)(w+1)(2w-1)}{35\pi^{2}w^{5}} \log(1-w)\bigg[ \frac{\left(12 z^4-24 z^3+z^2+46 z-23\right) z}{ (z-1)^3} \nonumber \\
\label{Kqf5by2}
& \hspace{2.5cm}-\frac{6 \left(2 z^2-7 z+7\right) z^5 }{ (1-z)^4} \log (z) + 6 (z-1) \left(2 z^2+3 z+2\right) \log (1-z)  \bigg].
\end{align} 




\subsection{Sums}\label{sums}

In this section, we will compute two types of sums that have appeared in the paper The first sum is encountered in the computation of discrete series kernel in section \ref{Discrete_Kernel} equation \eqref{eq:cl1}. The second sum is encountered in the computation of the kernel $K^{(p)}_{princ}(\Delta_{\phi},z,w)$ in appendix \ref{K_princ_p_Appen}, Eq.~\eqref{eq:plv2}.

\subsubsection*{Sum appeared in $K_{dis}(z,w)$ computation}

The sum to be computed in equation \eqref{eq:cl1} is as follows.
\begin{align}
\sum_{m \in 2 \mathbb{N}} (2m-1)Q_{m-1}(x) Q_{m-1}(y) \equiv \sum_{n \in \textrm{odd}} (2n+1)Q_{n}(x) Q_{n}(y) \ .
\end{align} 
To compute this sum, we need two ingredients, namely the orthogonality relation of LegendreP's and the relation between LegendreP's and LegendreQ's. These are given below.
\begin{align}
\sum_{n=0}^{\infty} (2n+1)P_{n}(x) P_{n}(y) &= 2 \delta(x-y) \ \forall \ x \in (-1,1) \qquad \textrm{(Orthogonality relation)} \label{orthoP} \\
Q_{n}(x) &= - \frac{1}{2} \int_{-1}^{1} dy  \frac{P_{n}(y)}{y-x} \ .\label{relationPQ}
\end{align}

\noindent Using the above two relations, we get
\begin{align}
\sum_{n=0}^{\infty} (2n+1)P_{n}(x) Q_{n}(y) &= - \int_{-1}^{1} dy'  \frac{\delta(x-y')}{y'-x} = \frac{1}{y-x} \ .
\end{align}

\noindent Now using the property of LegendreP's under parity i.e. $P_{n}(-x) = (-1)^n P_{n}(x)$, we get
\begin{align}
\sum_{n \in \textrm{odd}} (2n+1)P_{n}(x) Q_{n}(y) &= \frac{1}{2} \left( \sum_{n=0}^{\infty} (2n+1)P_{n}(x) Q_{n}(y) - \sum_{n=0}^{\infty} (2n+1)P_{n}(-x) Q_{n}(y) \right) \notag \\
& = \frac{1}{2} \left( \frac{1}{y-x} - \frac{1}{y+x}\right) \ .
\end{align}

\noindent Using the above equation along with the relation in Eq.~\eqref{relationPQ}, we get the result for the required sum.
\begin{align}
\sum_{n \in \textrm{odd}} (2n+1)Q_{n}(x) Q_{n}(y) &=  \frac{1}{2} \left(  - \frac{1}{2} \int_{-1}^{1} dx'  \frac{1}{(y-x')(x'-x)} + \frac{1}{2} \int_{-1}^{1} dx'  \frac{1}{(y+x')(x'-x)} \right) \notag \\
&= \frac{\log \left( \frac{1-x}{1+x} \frac{1+y}{1-y}\right)}{4(x-y)} + \frac{\log \left( \frac{1-x}{1+x} \frac{1-y}{1+y}\right)}{4(x+y)} \ .\label{sumQQ}
\end{align}

\subsubsection*{Sum appeared in $K^{(p)}_{princ}(\Delta_{\phi},z,w)$ computation}

The sum here to be computed in Eq.~\eqref{eq:plv2} is as follows.
\begin{align}
\sum_{m=0}^\infty   \frac{d}{dm'} \bigg[ (4m'+3)P_{2m'+1}(x) Q_{2m'+1}(y) \bigg] \bigg|_{m'=m} \equiv 2\sum_{m \in \textrm{odd}} \frac{d}{dm'} \Big[ (2 m'+1)P_{m'}(x)Q_{m' }(y) \Big]\bigg|_{m'=m}
\end{align} 

\noindent To compute the above sum we will use the following identity, 
\begin{align}
\sum_{m=0}^{\infty} \frac{d}{dm'} \Big[ (2 m'+1)P_{m'}(x)Q_{m' }(y) \Big]\bigg|_{m'=m}  &=  \frac{ 1 }{x-y} \frac{d}{d \a} \bigg[  \a\Big(   P_{ \a -1 }(x) Q_{ \a}(y) -P_{ \a}(x) Q_{ \a-1 }(y)  \Big)  \bigg]_{\a \to 0}   \notag\\
& =
 \frac{1}{2}\frac{1}{x-y} \log \Big[\frac{y+1}{x+1}\Big] \ .
\end{align}
where, in the second equality we have computed the rhs of the first line explicitly. Again using the property of LegendreP's under parity i.e. $P_{n}(-x) = (-1)^n P_{n}(x)$, we get
\begin{align}\label{sumdrPQ}
\sum_{m \in \textrm{odd}}^{\infty} \frac{d}{dm'} \Big[ (2 m'+1)P_{m'}(x)Q_{m' }(y) \Big]\bigg|_{m'=m} =
\frac{1}{2}\frac{1}{x-y} \log \Big[\frac{y+1}{x+1}\Big] + \frac{1}{2}\frac{1}{x+y} \log \Big[\frac{y-1}{x+1}\Big] \ .
\end{align}

\bibliographystyle{utphys}
\bibliography{1d_Draft1}

\end{document}